# Giant magnetoelectric effect induced by intrinsic surface stress in ferroic nanorods

M.D. Glinchuk,[1] E.A. Eliseev,[1] A.N. Morozovska,[1,*] and R.Blinc[2]

[1]Institute for Problems of Materials Science, NAS of Ukraine,
Krjijanovskogo 3, 03142 Kiev, Ukraine,

[2]Jozef Stefan Institute, P.O. Box 3000, 1001 Ljubljana, Slovenia

**Abstract**

The general approach for the consideration of the magnetoelectric effects in ferroic nanorods is proposed in the framework of the phenomenological theory. The intrinsic surface stress, magneto- and electrostriction as well as piezoelectric and piezomagnetic effects are included into the free energy. The intrinsic surface stress under the curved nanoparticle surface is shown to play an important role in the shift of ferroelectric and ferromagnetic transition temperatures and built-in magnetic and electric fields appearance, which are inversely proportional to the nanorod radius.

We consider the case of quadratic and linear magnetoelectric coupling coefficients. The linear coupling coefficient is radius independent, whereas the quadratic ones include terms inversely proportional to the nanorod radius and thus strongly increase with decrease of the radius. The predicted giant relative dielectric tunability in the vicinity of ferromagnetic and ferroelectric phase transition points induced by quadratic magnetoelectric coupling increases by 2-50 times. The quadratic magnetoelectric coupling dramatically changes the phase diagrams of ferroic nanorods when the radius decreases. In particular the second order phase transition may become a first one, the triple point state characterized by continuous set of order parameters appears at zero external electric and magnetic fields and the tricritical points appear under external fields.

Keywords: magnetoelectric effect, ferroic nanorods, surface stress.

PACS: 77.80.-e, 77.84.Dy, 68.03.Cd, 68.35.Gy

---

[*] Corresponding author: morozo@i.com.ua, permanent address: V. Lashkarev Institute of Semiconductor Physics, NAS of Ukraine, 41, pr. Nauki, 03028 Kiev, Ukraine

# Giant magnetoelectric effect induced by intrinsic surface stress in ferroic nanorods

**1. Introduction**

The magnetoelectric (ME) effect where the application of either a magnetic field or an electric field induces an electric polarization as well as magnetization attracted much attention in the last years. Although it was predicted by P. Curie [1] in 1894 on the basis of symmetry consideration and firstly observed in 1961 in an antiferromagnetic $Cr_2O_3$ crystal [2] the observed ME effect was small (about a few percents) that retarded its broad investigation and especially technical applications. The latter is related to the fact that ME materials must exhibit high ME coefficients for such important applications as magnetoelectric sensors in radioelectronics, optoelectronics, microwave electronics, transducers, magnetically tuned capacitors etc. Recently the revival of the ME effect has been observed due to discovery of high (several hundred percents) ME effects both in single phase and composite materials (see [3], [4, 5], [6] review [7] and ref. therein). Most of the composites exhibit a high extrinsic ME effect resulting from interaction between the magnetic and electric components via e.g. their magnetostriction and piezoelectric properties, as well as piezomagnetic-piezoelectric interaction [8, 9, 10]. The physical reason of the high ME effect is still unclear in single-phase materials, where the ME effect is intrinsic (see e.g. [11]). The description of this effect in microscopic theory based on a Hamiltonian with spin-orbit interaction frequently uses the idea that spin current symmetry belongs to the same class as the electric polarization and so it is natural to expect the coupling between them [12, 13]. Here however one could hardly expect a high ME coefficient.

The phenomenological theory approach of the description of the ME effect obligatory uses the interaction of magnetization and electric polarization with mechanical tension both in composites [8, 14] and single-phase materials [15, 16]. However no indication on the possibility to obtain high ME coupling was revealed for conventional type of mechanical conditions in the bulk materials.

It can be expected that because these conditions are completely different under the confinements of nanomaterials the probability to obtain high ME coupling would appear. Some evidence in favour of this supposition follows from the observation of dramatically higher ME coefficients in epitaxially (001) oriented $BiFeO_3$ films on a $SrTiO_3$ substrate than in the bulk crystals. This effect was suggested to be related to the influence of boundary conditions in the consideration performed in [17]. The authors of [17] came to the conclusion that the ME effect and other properties might be understood in terms of the appearance of a homogeneously magnetized state in the film. The same striking phenomenon was reported recently in the paper [18] about the observation of ferromagnetism in spherical nanoparticles (size 7–30 nm) of

Giant magnetoelectric effect induced by intrinsic surface stress in ferroic nanorods nonmagnetic oxides such as $CeO_2$, $Al_2O_3$, ZnO etc. Extremely strong superparamagnetic behaviour down to 4 K has been found in gold and palladium nanoparticles with mean diameter 2,5 nm with no magnetization in bulk and narrow sizes distribution [19]. The appearance of ferroelectricity was shown to take place in nanorods and films of the incipient ferroelectrics which stay paraelectric up to zero K in bulk [20, 21]. In these papers it was shown that the possible physical origin of electrical polarization and magnetization can be due to mechanical conditions in restricted geometry of nanomaterials and the influence of surface tension in particular. Therefore it is not excluded that new secondary ferroics with **M** and **P** might appear among oxide nanomaterials. This type of secondary ferroics is known to be a possibility for obtaining high ME effects [7]. However up to now no calculations were performed to find out if the restricted geometry and related mechanical boundary conditions could influence the value of ME effect coefficients.

In this paper we performed such calculations for the first time. We applied a phenomenological theory approach for oxide materials in the form of nanorods. Below one can see the details of the used model.

## 2. Theoretical approach
### *2.1. Model of calculations*
We will consider the secondary ferroic with two order parameters, magnetization **M** and electric polarization **P** in the form of nanorods. These order parameters can be either inherent in the bulk material or induced by confinements of nanorods. Keeping in mind that ferromagnetism has been observed at room temperature in nanoparticles of 7–30 nm size [18] and the sizes about 50 nm are suitable for the appearance of ferroelectricity [20] we suppose to consider nanorods with the sizes 5–50 nm. For such small sizes the influence of surface and related boundary conditions including surface tension are known to be high. Thus the properties are expected to be more close to those near the surface than in the bulk. While for larger sizes e.g. more than 100 nm the properties are known to change gradually from those on the surface to those in the bulk (see e.g. [22, 23]) for the considered sizes less than 50 nm the properties can be supposed to be homogeneous and under the strong influence of surface tension. The main mechanism of the mechanical tension relaxation is known to misfit dislocation. But the considered small sizes which are usually smaller than the critical size $\Delta h_d$ of dislocation appearance [24], the tension and thus the properties can be considered as homogeneous. As a matter of fact such an approach is in agreement with so called shell and core model of the nanoparticles [25] where the core is the inner part of a particle, that contrary to shell (outer part) does not "feel" the influence of

Giant magnetoelectric effect induced by intrinsic surface stress in ferroic nanorods surface. The core properties are thus practically the same as those in the bulk. Investigation of ferroelectric nanoparticles by ESR method had shown [26] that the shell sizes are in the region of several to tens nm. The characteristic feature of the shell has to be the absence of spatial inversion symmetry and so the existence of piezoelectric effect even for cubic symmetry in the bulk. Under the condition of appearance of ferromagnetism in the small nanoparticles [18] one can suggest the disappearance of time inversion symmetry and so piezomagnetic effect existence in nanoparticles. In what follows we will consider long nanosize cylindrical nanorods with electrical polarization along z axis and magnetization along one of three equivalent axes with the external electric and magnetic fields along z and x directions correspondingly (see Fig. 1). The nanorods are supposed to be clamped and long enough ($h >> R$). For the considered geometry the depolarization field is absent. Also it is possible to make the demagnetization field negligible [27]. Under such conditions single domain state will be the most preferable. The electro- and magnetostriction effects as well as mechanical stress tensor with boundary conditions at the curved nanoparticle surface must be taken into account. The nanorods are supposed to be separated from one another and so they are not interacting.

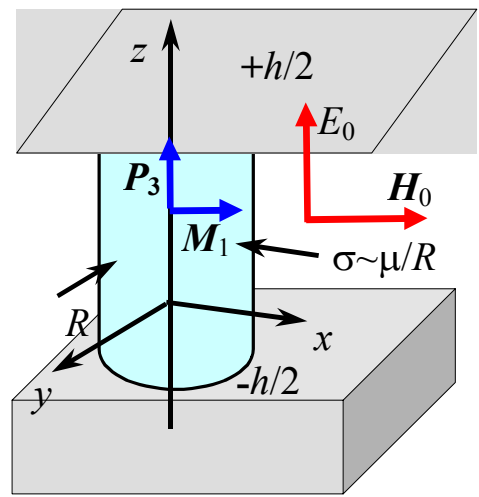

Fig.1. (Color online) Geometry of cylindrical particle, $x$ is one of the three equivalent weak magnetic anisotropy axes; $z$ is the polar ferroelectric axis. The external electric field $E_0$ is directed along polar axes, magnetic field $H_0$ is directed along the $x$ axes. The geometry $H_0 \perp E_0$ is typical for the majority of experiments.

The text of the paper contains several chapters which include the most general form of free energy $M$, $P$, striction and piezocoefficients and ME linear and nonlinear interaction are

Giant magnetoelectric effect induced by intrinsic surface stress in ferroic nanorods obtained after minimization of free energy with respect to stress components (section 2.2). Sections 3-4 are devoted to built-in electric and magnetic field and ME coupling coefficients with size effect calculations. The consideration of electric, magnetic and magnetoelectric susceptibility one can find in section 5. In section 6 the calculations of the transition temperature with and without ME interaction are presented. The size effects in the phase diagrams are considered too. In the section we consider the most widely spread case of existence of the squared ME effect only to show that even in this case size effects become very strong.

*2.2. Gibbs energy renormalization caused by intrinsic surface stresses*

Let us consider a rather long cylindrical ferroic nanorod with free sidewalls ($\rho = R$) clamped between the wafer ($z = -h/2$) and top electrode ($z = +h/2$) (see Fig.1).

In what follows we will consider one set of parameters for the nanoparticle. Such a situation is possible for a nanorod of small radius. However it is not excluded that the numerical values and symmetry of material tensorial constants differ from the ones tabulated for bulk material, e.g. there are "shell" constants inter-grown through the nanoparticle "core". For this important case several electro- and/or magneto-mechanical coupling phenomena absent in the bulk may appear from the symmetry breaking in the vicinity of surface (see below).

Let us suppose that polarization is directed along the polar axis z and magnetization along weak magnetic anisotropy axis x, i.e. $\mathbf{P} = (0, 0, P_3)$ and $\mathbf{M} = (M_1, 0, 0)$. So, the Gibbs energy expansion of the homogeneous polarization $P_3(T)$, the magnetization component $M_1(T)$ and the stress $\sigma_{ij}$ has the form:

$$G_R = 2\pi h \int_0^R \rho d\rho \begin{pmatrix} a_1 P_3^2 + a_{11} P_3^4 + a_{111} P_3^6 - (Q_{11}\sigma_{33} + Q_{12}(\sigma_{11} + \sigma_{22}))P_3^2 + \\ -\frac{1}{2}(A_{11}\sigma_{11}^2 + A_{11}\sigma_{22}^2 + A_{33}\sigma_{33}^2)P_3^2 + ... - g_{3jk}^e \sigma_{jk} P_3 + \\ b_1 M_1^2 + b_{11} M_1^4 + a_{111} M_1^6 - (Z_{11}\sigma_{33} + Z_{12}(\sigma_{11} + \sigma_{22}))M_1^2 - \\ -\frac{1}{2}(B_{11}\sigma_{11}^2 + B_{11}\sigma_{22}^2 + B_{33}\sigma_{33}^2)M_1^2 + ... - g_{1jk}^m \sigma_{jk} M_1 - \\ -\frac{1}{2}s_{11}(\sigma_{11}^2 + \sigma_{22}^2 + \sigma_{33}^2) - s_{12}(\sigma_{11}\sigma_{22} + \sigma_{11}\sigma_{33} + \sigma_{33}\sigma_{22}) - \\ -\frac{1}{2}s_{44}(\sigma_{23}^2 + \sigma_{13}^2 + \sigma_{12}^2) + f_{ijkl}\sigma_{ij}^2 \sigma_{kl}^2 - M_1 H_0 - P_3 E_0 \end{pmatrix} \quad (1)$$

Subscripts 1, 2 and 3 denote Cartesian coordinates x, y and z respectively, we use Voigt notation or matrix notation when necessary (xx=1, yy=2, zz=3, zy=4, zx=5, xy=6). Coefficients $a_1(T)$ and $b_1(T)$ explicitly depend on temperature $T$ in the framework of the Landau-Ginzburg

Giant magnetoelectric effect induced by intrinsic surface stress in ferroic nanorods approach. All the higher order expansion coefficients are supposed to be temperature independent. Since we supposed that the order parameters and elastic stress spatial distribution are homogeneous inside the nanorod, we should neglect the gradient energy. Note that for the film on the substrate this supposition is valid when the film thickness is less than the critical thickness of misfit dislocation appearance that is known to be dozens of nm [24].

In Eq.(1) $Q_{ij}$ and $Z_{ij}$ are respectively the electro- and magneto-striction tensor coefficients; $s_{ij}$ are components of elastic compliance tensor [28]. Hereinafter we suppose the symmetry of piezoelectric ($g^e_{3jk}$) and piezomagnetic ($g^m_{3jk}$) effects due to the surface influence are different from the cubic phase as follows: $g^e_{3jk}\sigma_{jk}P_3 = g^e_{31}(\sigma_{11}+\sigma_{22})P_3 + g^e_{33}\sigma_{33}P_3 + ...$ and $g^m_{1jk}\sigma_{jk}M_1 = g^m_{11}(\sigma_{11}+\sigma_{22})M_1 + g^m_{13}\sigma_{33}M_1 + ......$

The distribution of stress $\sigma_{ij}$ should satisfy the conditions of mechanical equilibrium as well as appropriate boundary conditions at the curved nanoparticle surface:

$$\begin{cases} \dfrac{\partial \sigma_{ij}}{\partial x_i} = 0, \\ \sigma_{\rho\rho}\big|_{\rho=R} = -\dfrac{\mu}{R}, \quad \sigma_{\rho\varphi}\big|_{\rho=R} = 0, \quad \sigma_{\rho z}\big|_{\rho=R} = 0, \quad u_{33}(z=\pm h/2) = 0 \end{cases} \quad (2)$$

Here $\mu_{ij} = \mu\delta_{ij}$ is the intrinsic surface stress tensor coefficients [29, 30] which have nontrivial components only on the nanorod surface. The surface stress $\mu$ is strongly dependent on the ambient material.

The minimization of the free energy (1) with respect to the stress components $\sigma_{ij}$ leads to the equations of state $\partial G_R/\partial \sigma_{ij} = -u_{ij}$. Neglecting the terms $\sim f_{ijkl}\sigma^2_{ij}\sigma^2_{kl}$ one obtains:

$$\begin{cases} s_{11}\sigma_{11} + s_{12}(\sigma_{22}+\sigma_{33}) + (Q_{12}+A_{11}\sigma_{11})P_3^2 + (Z_{12}+B_{11}\sigma_{11})M_1^2 + g^e_{31}P_3 + g^m_{11}M_1 = u_{11}, \\ s_{11}\sigma_{22} + s_{12}(\sigma_{11}+\sigma_{33}) + (Q_{12}+A_{11}\sigma_{22})P_3^2 + (Z_{12}+B_{11}\sigma_{22})M_1^2 + g^e_{31}P_3 + g^m_{11}M_1 = u_{22}, \\ s_{11}\sigma_{33} + s_{12}(\sigma_{22}+\sigma_{11}) + (Q_{11}+A_{33}\sigma_{33})P_3^2 + (Z_{11}+B_{33}\sigma_{33})M_1^2 + g^e_{33}P_3 + g^m_{13}M_1 = u_{33}, \\ 2u_{23} = s_{44}\sigma_{23}, \quad 2u_{13} = s_{44}\sigma_{13}, \quad 2u_{12} = s_{44}\sigma_{12}. \end{cases} \quad (3)$$

Here $u_{ij} = (\partial u_i/\partial x_j + \partial u_j/\partial x_i)/2$ are strain tensor components ($u_i$ is the displacement vector components).

The homogeneous solution of Eqs.(3) for the stress and strain tensor components $\sigma_{ij}$ in Cartesian coordinates has the form:

$$\sigma_{11} = \sigma_{22} = -\frac{\mu}{R}, \quad \sigma_{12} = \sigma_{13} = \sigma_{23} = 0, \quad (4a)$$

Giant magnetoelectric effect induced by intrinsic surface stress in ferroic nanorods

$$\sigma_{33} = \frac{s_{12}(2\mu/R) - Q_{11}P_3^2 - Z_{11}M_1^2 - g_{33}^e P_3 - g_{13}^m M_1}{s_{11} + A_{33}P_3^2 + B_{33}M_3^2} \approx$$

$$\approx \left(\frac{s_{12}}{s_{11}}\frac{2\mu}{R} - \frac{Q_{11}P_3^2 + Z_{11}M_1^2 + g_{33}^e P_3 + g_{13}^m M_1}{s_{11}}\right)\left(1 - \frac{A_{33}P_3^2 + B_{33}M_1^2}{s_{11}}\right) \quad (4b)$$

$$u_{23} = u_{13} = u_{12} = u_{33} = 0, \quad (4c)$$

$$u_{11} = u_{22} = \left(\begin{array}{l} s_{12}\dfrac{s_{12}(2\mu/R) - Q_{11}P_3^2 - Z_{11}M_1^2 - g_{33}^e P_3 - g_{13}^m M_1}{s_{11} + A_{33}P_3^2 + B_{33}M_1^2} + \\ + Q_{12}P_3^2 + Z_{12}M_1^2 + g_{31}^e P_3 + g_{11}^m M_1 - \dfrac{\mu}{R}\left(s_{11} + s_{12} + A_{11}P_3^2 + B_{11}M_1^2\right) \end{array}\right) \quad (4d)$$

Rigorously speaking, solutions (4) are valid for nanorods of radius $R$ less that the critical thickness $\Delta h_d$ of surface stress relaxation (e.g. dislocation appearance [24]). For the case $R \gtrsim \Delta h_d$ a rather complex inhomogeneous elastic problem with elastic stress and polarization gradients should be considered, that is far beyond the scope of the paper. Approximate solution could be obtained within the framework of conventional core and shell model [23]. Elastic stress $\sigma_{ij}$ given by Eqs. (4a,b) is mainly concentrated in the shell region, whereas the core is almost unstressed (i.e. $\sigma_{ij} \approx 0$).

Typically $\Delta h_d \sim 5 - 50$ nm. We thus mainly consider the case $R < \Delta h_d$ (i.e. all the particles are in shell in sections 3-6), since it is the most interesting one for surface and size effects manifestation. Effects related with the shell influence on the dielectric and magnetoelectric properties of thick nanorods will be qualitatively considered in the Discussion.

Hereinafter we suppose that the terms $A_{ii}\sigma_{ii}^2 P_3^2$ and $B_{ii}\sigma_{ii}^2 M_1^2$ are small, so we neglect their higher powers. Substituting Eq.(4a-b) into Eq. (1) we obtain the Gibbs energy with renormalized coefficients:

$$G_R = 2\pi h \int_0^R \rho\, d\rho \left(\begin{array}{l} \alpha_1(T,R)P_3^2 + \alpha_{11}P_3^4 - P_3\left(E_0 + E_p(R)\right) \\ + \beta_1(T,R)M_1^2 + \beta_{11}M_1^4 + M_1\left(H_0 + H_p(R)\right) + g_{ME}(P_3, M_1) \end{array}\right). \quad (5)$$

The renormalized coefficients before $P_3^2$ and $M_1^2$ in the free energy (5) have the form:

$$\alpha_1(T,R) = a_1(T) + \frac{\left(g_{33}^e\right)^2}{2s_{11}} + \frac{2\mu}{R}\left(Q_{12} - Q_{11}\frac{s_{12}}{s_{11}}\right) - \frac{2\mu^2}{R^2}\left(A_{11} + A_{33}\frac{s_{12}^2}{s_{11}^2}\right), \quad (6a)$$

$$\beta_1(T,R) = b_1(T) + \frac{\left(g_{13}^m\right)^2}{2s_{11}} + \frac{2\mu}{R}\left(Z_{12} - Z_{11}\frac{s_{12}}{s_{11}}\right) - \frac{2\mu^2}{R^2}\left(B_{11} + B_{33}\frac{s_{12}^2}{s_{11}^2}\right). \quad (6b)$$

Giant magnetoelectric effect induced by intrinsic surface stress in ferroic nanorods

The internal "built-in" fields induced by the piezoelectric and piezomagnetic effects are introduced as

$$E_p(R) = \left(\frac{s_{12}}{s_{11}} g^e_{33} - g^e_{31}\right)\frac{4\mu}{R}, \tag{7a}$$

$$H_p(R) = \left(\frac{s_{12}}{s_{11}} g^m_{13} - g^m_{11}\right)\frac{4\mu}{R}. \tag{7b}$$

The nanorod (i.e. shell) magnetoelectric energy density is introduced as:

$$g_{ME} = \left(\gamma_{11} M_1 P_3 + \gamma_{12} M_1 P_3^2 + \gamma_{21} M_1^2 P_3 + \gamma_{22} M_1^2 P_3^2\right) \tag{8}$$

Linear and quadratic magnetoelectric coupling coefficients in the magnetoelectric energy (8) are:

$$\gamma_{11} = \frac{g^e_{33} g^m_{13}}{s_{11}}, \tag{9a}$$

$$\gamma_{12} = g^m_{13}\left(\frac{Q_{11}}{s_{11}} + \frac{2\mu}{R}\frac{s_{12} A_{33}}{s_{11}^2}\right), \quad \gamma_{21} = g^e_{33}\left(\frac{Z_{11}}{s_{11}} + \frac{2\mu}{R}\frac{s_{12} B_{33}}{s_{11}^2}\right), \tag{9b}$$

$$\gamma_{22} = \left(\frac{Q_{11} Z_{11}}{s_{11}} - \frac{A_{33}\left(g^m_{13}\right)^2 + B_{33}\left(g^e_{33}\right)^2}{2 s_{11}^2} + \frac{2\mu}{R}\frac{s_{12}}{s_{11}^2}(Q_{11} B_{33} + Z_{11} A_{33}) + \frac{4\mu^2}{R^2}\frac{s_{12}^2}{s_{11}^3} B_{33} A_{33}\right) \tag{9c}$$

Both nonzero values $g^e_{33} \neq 0$ and $g^m_{13} \neq 0$ are necessary to obtain nonzero linear coupling coefficient $\gamma_{11} \neq 0$ that may be possible in some special cases. For instance, $g^m_{33} = 1.2 \cdot 10^{-8}$ Wb/N and $g^m_{13} = -5.8 \cdot 10^{-9}$ Wb/N in Terfenol-D [9], but piezoelectric coupling is absent in the bulk. However it may appear inside the nanorod of radius $R \leq \Delta h_d$ (or the corresponding shell region for $R > \Delta h_d$) allowing for the symmetry breaking on the surface. The coefficient $\gamma_{22} \neq 0$ is non-zero for all magnetoelectric materials. Note, that the last term in Eq.(9c) proportional to the product $B_{33} A_{33}$ would be neglected hereinafter.

For the case $R \gg \Delta h_d$ the core (i.e. bulk) magnetoelectric coupling coefficients are radius independent and the magnetoelectric energy coincides with the one of laterally clamped bulk material:

$$g_{ME} \approx \frac{g^e_{33} g^m_{13}}{s_{11}} M_1 P_3 + g^m_{13}\frac{Q_{11}}{s_{11}} M_1 P_3^2 + g^e_{33}\frac{Z_{11}}{s_{11}} M_1^2 P_3 + \frac{Q_{11} Z_{11}}{s_{11}} M_1^2 P_3^2 \tag{10}$$

### 3. Built-in fields thickness dependence

The built-in electric (7a) and magnetic (7b) fields dependence on nanorod radius are shown in Fig.2 for typical material parameters. It is clear from the log-log dependencies that built-in fields increase with a radius decrease and could overcome the bulk coercive field values $E_C$ and $H_C$ (see solid and dashed lines).

Built-in electric field leads to the horizontal shift of all hysteresis loops and electret-like state appearance in ferroelectric films with thickness less than the critical one. It facilitates thin film the self-polarization as predicted in Refs.[31, 32]. Beside trivial hysteresis loop horizontal shifts (see insets 2c,d), we predict ordering effects caused by built-in magnetic fields, radius dependent $H_p(R)$ effect may induce ferromagnetism or irreversible magnetization in small nanorods absent in the bulk material. This is similar to the ferroelectricity in incipient ferroelectric nanorods [20] and electret state in ultrathin films [32]. Under the absence of external magnetic field, the built-in magnetic field smears the magnetic, dielectric and the magnetoelectric susceptibility temperature maximum, increases their values in paramagnetic phase and essentially increases magnetoelectric tunability.

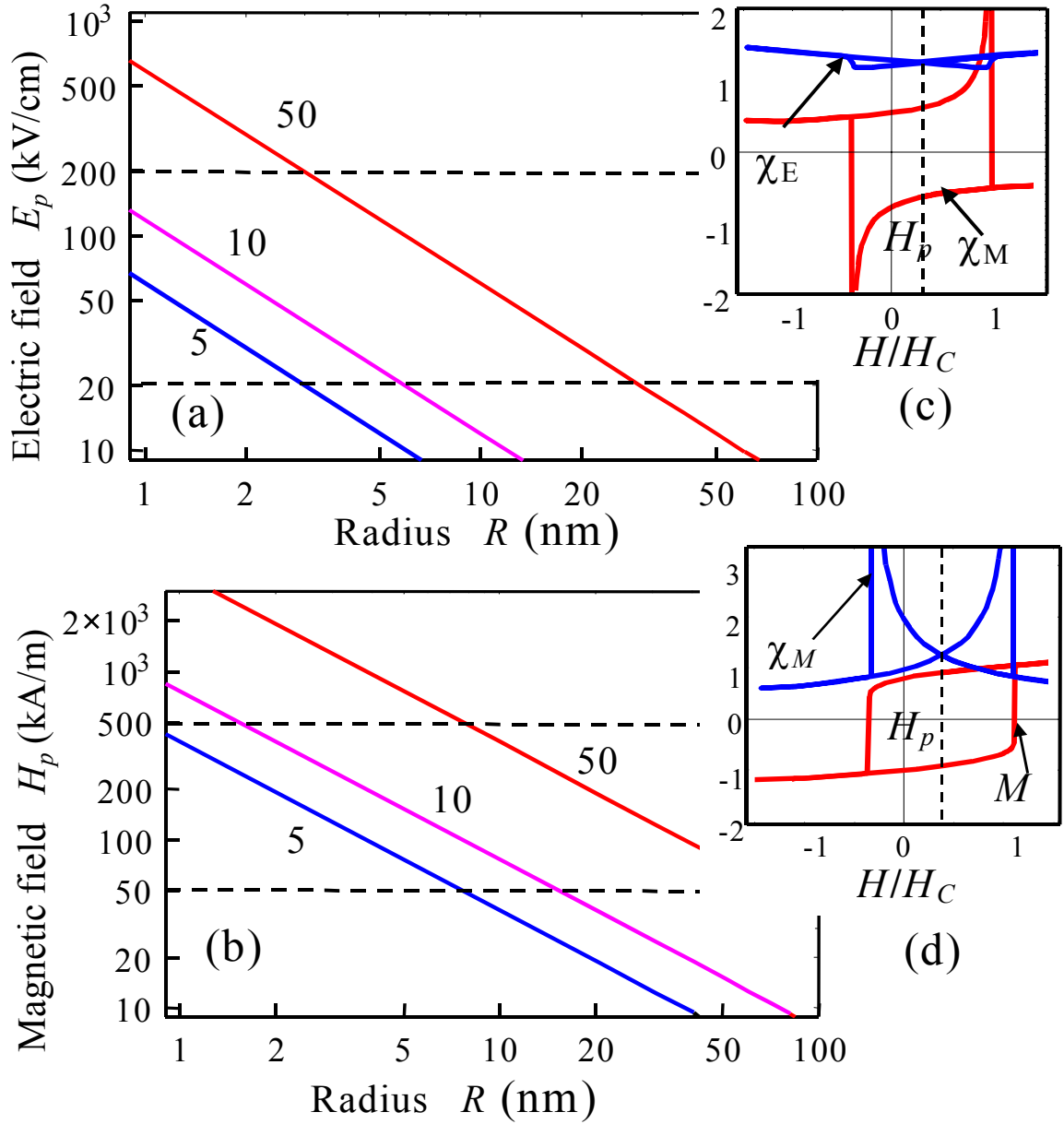

Fig.2. (Color online) Built-in electric (a) and magnetic (b) fields dependence on nanorod radius for parameters $g_{13}^m \cong 10^{-9}$ Wb/N, $g_{11}^m = 0$, $g_{33}^e \cong 10^{-3}$ Vm/N, $g_{31}^e = 0$, $\mu = 5$, 10, 50 N/m (figures near the curves) and $s_{12}/s_{11} = -0.3$. Region between the dashed lines corresponds to typical range of bulk coercive fields $E_C$ and $H_C$ (see e.g. [33]). Insets (c,d): schematic hysteresis loops via the normalized magnetic field $H/H_C$. All values are normalized on their bulk values at zero fields without electromagnetic coupling. The electric field is assumed to be zero.

Giant magnetoelectric effect induced by intrinsic surface stress in ferroic nanorods

**4. Size effect on magnetoelectric coupling coefficients**

Let us rewrite Eqs.(9) as

$$\gamma_{12}(R) = \gamma_{12}^b\left(1+\frac{R_{12}}{R}\right), \quad \gamma_{21}(R) = \gamma_{21}^b\left(1+\frac{R_{21}}{R}\right), \quad \gamma_{22}(R) \approx \gamma_{22}^b\left(1+\frac{R_{22}}{R}\right). \tag{11}$$

Where $\gamma_{12}^b = g_{13}^m \frac{Q_{11}}{s_{11}}$, $R_{12} = 2\mu \frac{s_{12} A_{33}}{s_{11} Q_{11}}$, $\gamma_{21}^b = g_{33}^e \frac{Z_{11}}{s_{11}}$, $R_{21} = 2\mu \frac{s_{12} B_{33}}{s_{11} Z_{11}}$,

$\gamma_{22}^b = \frac{Q_{11} Z_{11}}{s_{11}} - \frac{A_{33}(g_{13}^m)^2 + B_{33}(g_{33}^e)^2}{2 s_{11}^2}$ and $R_{22} = R_{12} + R_{21} = 2\mu \frac{s_{12}(Q_{11} B_{33} + Z_{11} A_{33})}{s_{11} Q_{11} Z_{11}}$. Usually bulk material magnetoelectric coupling constants $\gamma_{ij}^b$ are small or identically zero depending on material symmetry.

The linear coupling coefficient is radius independent, whereas the quadratic ones include terms inversely proportional to the nanorod radius. They thus strongly increase with radius decrease. Linear magnetoelectric coupling $\gamma_{11}$ breaks the symmetry $P \to -P$ and $M \to -M$ as well as smears the transition point even at zero magnetic and electric fields. So, Eq.(11) allows the strong renormalization and even sign change of $\gamma_{ij \neq 11}$ caused by intrinsic surface stress, since characteristic parameters $R_{ij}$ could be positive or negative. In accordance with estimations made in Appendix A, usually $1\,\text{nm} \leq |R_{12}| \leq 50\,\text{nm}$, $5\,\text{nm} \leq |R_{21}| \leq 50\,\text{nm}$ and so their sum $|R_{22}| \leq 100\,\text{nm}$. Size dependence of the normalized coupling coefficients is shown in Fig.3. It is clear from Fig.3, that at small radiuses $R/|R_{ij}| \ll 1$ coefficients $\gamma_{ij \neq 11}$ are much greater than their bulk values $\gamma_{ij \neq 11}^b$.

Giant magnetoelectric effect induced by intrinsic surface stress in ferroic nanorods

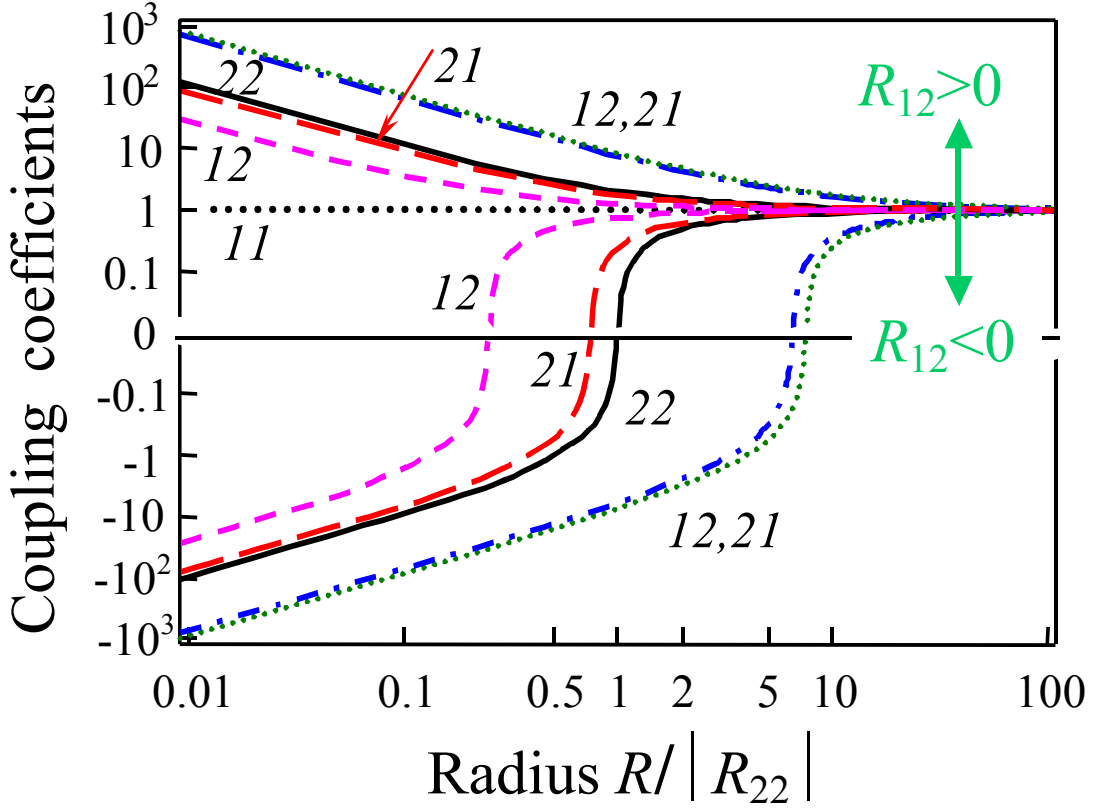

Fig.3. (Color online) Size dependence of the normalized coupling coefficients $\tilde{\gamma}_{ij} = \gamma_{ij}/|\gamma^b_{ij}|$ via the nanorod radius $R/|R_{22}|$: $\tilde{\gamma}_{22}$ (solid curves "22"), $\tilde{\gamma}_{12}$ ("12" long-dashed curves for $R_{12}/|R_{22}| = \pm 0.75$ and dotted curves for $R_{12}/|R_{22}| = \pm 7.5$), $\tilde{\gamma}_{12}$ ("21" short-dashed curves for $R_{12}/|R_{22}| = \pm 0.25$ and dash-dotted curves for $R_{12}/|R_{22}| = \pm 6.5$) and constant $\tilde{\gamma}_{11} = 1$ ("11" circles). To generate the plots we used the identity $R_{21}/|R_{22}| = 1 - R_{12}/|R_{22}|$.

Below we demonstrate the effects related with magnetoelectric coupling coefficients $\gamma_{ij \neq 11}$ and their influence on ferromagnetic and ferroelectric transition temperatures. We show the increase of the ferroelectric and ferromagnetic transition temperatures with the decrease of the nanorods radius in comparison with the bulk material.

## 5. Generalized susceptibilities calculations

Using renormalized coefficients (6, 7, 9), one can rewrite the free energy (5, 8) density $\tilde{g}(R,T)$ as follows:

Giant magnetoelectric effect induced by intrinsic surface stress in ferroic nanorods

$$\widetilde{g}(R,T) = \begin{pmatrix} \alpha_1 P_3^2 + \beta_1 M_1^2 + \alpha_{11} P_3^4 + \beta_{11} M_1^4 - (E_p + E_0)P_3 - (H_p + H_0)M_1 + \\ + \gamma_{11} M_1 P_3 + \gamma_{12} M_1 P_3^2 + \gamma_{21} M_1^2 P_3 + \gamma_{22} M_1^2 P_3^2 \end{pmatrix} \quad (12)$$

The conditions of the free energy minimum $\partial \widetilde{g}/\partial P_3 = 0$ and $\partial \widetilde{g}/\partial M_1 = 0$ lead to the coupled equations of state:

$$\begin{cases} 2\alpha_1 P_3 + \gamma_{11} M_1 + 2\gamma_{12} M_1 P_3 + \gamma_{21} M_1^2 + 2\gamma_{22} P_3 M_1^2 + 4\alpha_{11} P_3^3 = E_p + E_0 \\ 2\beta_1 M_1 + \gamma_{11} P_3 + \gamma_{12} P_3^2 + 2\gamma_{21} M_1 P_3 + 2\gamma_{22} P_3^2 M_1 + 4\beta_{11} M_1^3 = H_p + H_0 \end{cases} \quad (13)$$

After elementary transformations listed in Appendix B, the electric and magnetic susceptibilities can be found from the system (14) along with the magnetoelectric coupling coefficient:

$$\chi_E = \frac{\partial P_3}{\partial E_0} = \frac{2(\beta_1 + \gamma_{21} P_3 + \gamma_{22} P_3^2 + 6\beta_{11} M_1^2)}{\Delta(M_1, P_3)}, \quad (14a)$$

$$\chi_M = \frac{\partial M_1}{\partial H_0} = \frac{2(\alpha_1 + \gamma_{12} M_1 + \gamma_{22} M_1^2 + 6\alpha_{11} P_3^2)}{\Delta(M_1, P_3)}, \quad (14b)$$

$$\chi_{ME} = \frac{\partial P_3}{\partial H_0} = \frac{\partial M_1}{\partial E_0} = -\frac{\gamma_{11} + 2\gamma_{12} P_3 + 2\gamma_{21} M_1 + 4\gamma_{22} M_1 P_3}{\Delta(M_1, P_3)} \quad (14c)$$

Where:

$$\Delta(M_1, P_3) = \begin{pmatrix} 4(\alpha_1 + \gamma_{12} M_1 + \gamma_{22} M_1^2 + 6\alpha_{11} P_3^2)(\beta_1 + \gamma_{21} P_3 + \gamma_{22} P_3^2 + 6\beta_{11} M_1^2) - \\ -(\gamma_{11} + 2\gamma_{12} P_3 + 2\gamma_{21} M_1 + 4\gamma_{22} M_1 P_3)^2 \end{pmatrix} \quad (15)$$

Built-in fields (7) lead to the order parameter and susceptibility hysteresis loop horizontal shift, i.e. the loop horizontal scale is determined by the sum $H_p(R) + H_0$ as shown in Figs.2c,d.

The temperature dependence of susceptibilities $\chi_E$ and $\chi_{ME}$ are presented in Figs.4a-b at zero electric field and increasing magnetic field $H_0$ for positive coupling coefficient $\gamma_{22}(R) > 0$, zero coefficients $\gamma_{21} = \gamma_{12} = \gamma_{11} = 0$ and neglecting size effects of the transition temperatures. The corresponding ferroelectric order parameter $P_3$ is shown in Fig.4c. The temperature dependence of dielectric tunability $\delta \chi_E = (\chi_E(H) - \chi_E(0))/\chi_E(0)$ for zero electric field $E = 0$ and different values of magnetic field $H_0$ (in coercive field units) is shown in Fig.4d.

When generate the plots we put coefficients $\alpha_1 \cong \alpha_T (T - T_{ce}^b)$, $\beta_1 \cong \beta_T (T - T_{cm}^b)$, where $T_{ce}^b$ and $T_{cm}^b$ are the bulk ferroelectric and ferromagnetic transition temperatures. Note that the condition $\gamma_{21} = \gamma_{12} = 0$ leads to zero built-in fields in accordance with Eqs.(9).

Giant magnetoelectric effect induced by intrinsic surface stress in ferroic nanorods

Detailed consideration of the size-induced transition temperature shift will be presented in the next section. Here we demonstrate that under the condition $\gamma_{22}(R) \gg |\gamma_{22}^b|$ the influence of the quadratic ME coupling term $\gamma_{22} M_1^2 P_3^2$ on susceptibilities and order parameters may be strong itself.

It is clear from Fig.4a that the dielectric inverse susceptibility $\chi_E^{-1}$ decreases (so its direct value $\chi_E$ increases) and smears with $H_0$ increase. At relatively small magnetic fields $H_0/H_C \ll 1$ the susceptibility jump-like peculiarity appears at temperature lower than the bulk ferromagnetic phase transition temperature $T_{cm}^b$ (see curve 1). The jump diffuses and shifts to higher temperatures with magnetic field strength increase (see curves 2-3). The shift from the bulk transition temperature $T_{cm}^b$ depends on the ME coupling as $\beta_1 + \gamma_{22} P_3^2$ (see also Eq.(14-15)). So, ME coupling increases the ferromagnetic transition temperature at negative $\gamma_{22}$ and decreases it at positive $\gamma_{22}$. We predict that high enough ME coupling ($\gamma_{22}(R) < 0$, $|\gamma_{22}(R)| \gg \gamma_{22}^b$) may lead to the condition $\beta_1(T) + \gamma_{22} P_3^2 \leq 0$ in some temperature range even at $\beta_1(T=0) \geq 0$ and thus induces a ferromagnetic phase in small nanorods. It is absent in the bulk material (similarly to the appearance of ferroelectricity in incipient ferroelectric nanorods [20]).

At high magnetic fields $H_0/H_C > 1$ the ME coupling induces ferroelectric-paraelectric phase transition at low temperatures $T \ll T_{ce}^b$ (see the point $\chi_E^{-1} = 0$ at curve 4 in Fig.4a). ME coupling-induced ferroelectric phase transition appeared under the condition $\alpha_1 + \gamma_{22} M_1^2 \geq 0$, i.e. when positive coupling term $\gamma_{22} M_1^2 P_3^2$ suppresses the ferroelectric phase as shown in Fig.4c for the polarization $P_3$. No such ferroelectric-paraelectric phase transition is observed at negative $\gamma_{22}$, in contrast, the bulk phase transition $T = T_{ce}^b$ moves to the higher temperatures.

It is clear from Fig.4b that magnetolectric susceptibility $\chi_{ME}$ is almost absent in paramagnetic phase at small magnetic field. Its maximum diffuses with increasing magnetic field strength. Thus the $\chi_{ME}$ temperature behavior is similar to the pyroelectric coefficient one.

It is clear from Fig.4d, that the dielectric tunability increases under the magnetic field increase. The jump on the tunability appeared at lower temperatures than $T_{cm}^b$. It is related with the ferromagnetic phase transition shifted by the ME coupling. The jump height increases with magnetic field increase. Giant tunability appearance at high enough magnetic fields $H_0/H_C > 1$

Giant magnetoelectric effect induced by intrinsic surface stress in ferroic nanorods

(see divergence on curve 4 in Figs.4d) is caused by the ME coupling-induced ferroelectric-paraelectric phase transition taken place at positive $\gamma_{22}$ values (see the point, where $\chi_E^{-1} = 0$ at curve 4 for in Figs.4a and $P_3 = 0$ in Fig.4c). The dielectric tunability is colossal (2-50 times) in the vicinity of the phase transition (compare with 500% effect shown in Fig.2 from Ref.[11]). The size effects are absent in the bulk material allowing for negligibly small bulk ME coupling coefficients.

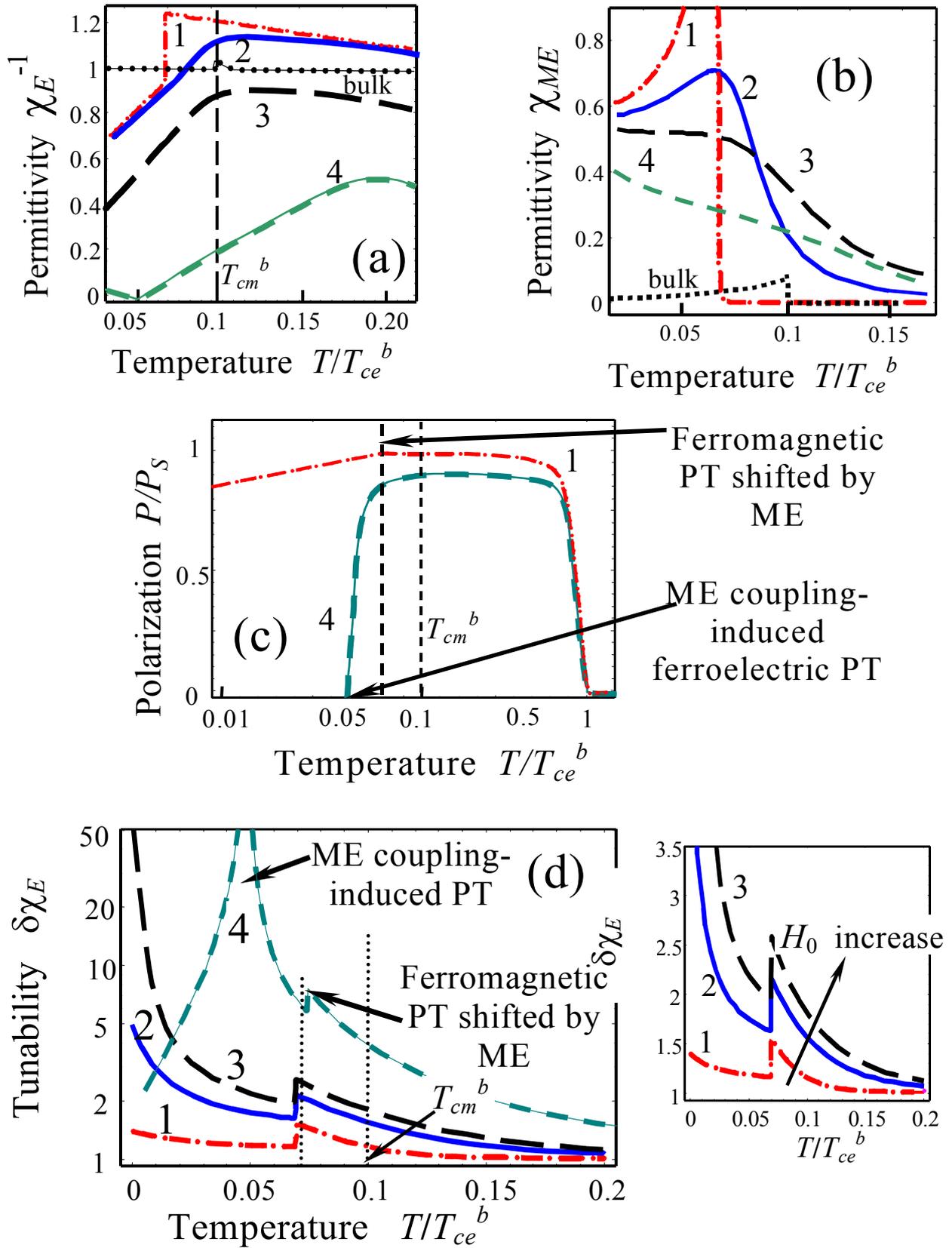

Fig.4. (Color online) Temperature dependence of dielectric susceptibility $\chi_E$ (a), magnetoelectric susceptibility $\chi_{ME}$ (b), polarization $P_3$ (c) and dielectric tunability $\delta\chi_E$ (d) at zero electric field $\gamma_{21} = \gamma_{12} = \gamma_{11} = 0$, $\gamma_{22}/|\gamma_{22}^b| = 50$, $\beta_1(0)/\sqrt{\beta_{11}} = 1.5\alpha_1(0)/\sqrt{\alpha_{11}}$, bulk ferromagnetic to ferroelectric transition temperatures ratio $T_{cm}^b/T_{ce}^b = 0.1$ and increasing values of magnetic field $H_0/H_C = 0.01; 0.5; 1; 2$ (curves 1, 2, 3, 4). The values $\chi_E$ and $\chi_{ME}$ are compared with corresponding bulk values (dotted curves) at zero fields and typical electromagnetic coupling.

## 6. Size effect on phase diagrams

*6.1. Size effect on transition temperatures without ME coupling*

The coefficients $\alpha_1$ and $\beta_1$ are temperature and radius dependent in accordance with Eqs.(6a,b). They can be rewritten as $\alpha_1 = \alpha_T(T - T_{ce}(R))$, $\beta_1 = \beta_T(T - T_{cm}(R))$. The temperatures $T_{ce}(R)$ and $T_{cm}(R)$ are:

$$\begin{cases} T_{ce}(R) = T_{ce}^*\left(1 - \dfrac{R_Q}{R} + \dfrac{\rho_A}{R^2}\right), & T_{ce}^* = T_{ce}^b + \dfrac{(g_{33}^e)^2}{2s_{11}\alpha_T}, \\ R_Q = \dfrac{2\mu}{\alpha_T T_{ce}^*}\left(Q_{12} - Q_{11}\dfrac{s_{12}}{s_{11}}\right), & \rho_A = \dfrac{2\mu^2}{\alpha_T T_{ce}^*}\left(A_{11} + A_{33}\dfrac{s_{12}^2}{s_{11}^2}\right) \end{cases} \quad (16a)$$

$$\begin{cases} T_{cm}(R) = T_{cm}^*\left(1 - \dfrac{R_Z}{R} + \dfrac{\rho_B}{R^2}\right), & T_{cm}^* = T_{cm}^b + \dfrac{(g_{13}^m)^2}{2s_{11}\beta_T}, \\ R_Z = \dfrac{2\mu}{\beta_T T_{cm}^*}\left(Z_{12} - Z_{11}\dfrac{s_{12}}{s_{11}}\right), & \rho_B = \dfrac{2\mu^2}{\beta_T T_{cm}^*}\left(B_{11} + B_{33}\dfrac{s_{12}^2}{s_{11}^2}\right) \end{cases} \quad (16b)$$

In order to consider the case of $T_{cm}^b = 0$, $T_{ce}^b = 0$ (i.e. when bulk material has no ferroic properties), one should use the following dependencies:

$$\begin{cases} T_{ce}(R) = T_{ce}^* - \dfrac{1}{\alpha_T}\left(\dfrac{R_Q}{R} - \dfrac{\rho_A}{R^2}\right), & T_{ce}^* = T_{ce}^b + \dfrac{(g_{33}^e)^2}{2s_{11}\alpha_T}, \\ R_Q = 2\mu\left(Q_{12} - Q_{11}\dfrac{s_{12}}{s_{11}}\right), & \rho_A = 2\mu^2\left(A_{11} + A_{33}\dfrac{s_{12}^2}{s_{11}^2}\right) \end{cases} \quad (16c)$$

$$\begin{cases} T_{cm}(R) = T_{cm}^* - \dfrac{1}{\beta_T}\left(\dfrac{R_Z}{R} - \dfrac{\rho_B}{R^2}\right), & T_{cm}^* = T_{cm}^b + \dfrac{(g_{13}^m)^2}{2s_{11}\beta_T}, \\ R_Z = 2\mu\left(Z_{12} - Z_{11}\dfrac{s_{12}}{s_{11}}\right), & \rho_B = 2\mu^2\left(B_{11} + B_{33}\dfrac{s_{12}^2}{s_{11}^2}\right) \end{cases} \quad (16d)$$

Giant magnetoelectric effect induced by intrinsic surface stress in ferroic nanorods

The temperatures $T_{ce}(R)$ and $T_{cm}(R)$ determine the points of corresponding paraphase instability. Under the negligibly small magnetoelectric energy (8), $T_{ce}(R)$ and $T_{cm}(R)$ determine the second order ferroelectric and ferromagnetic phase transition points correspondingly.

Schematic dependence of the temperatures $T_{ce}(R)$ and $T_{cm}(R)$ via the nanorod radius $R$ is shown in Fig.5. Note, that both signs of characteristic constants $R_{Q,Z}$ and $\rho_{A,B}$ are possible. Estimations [20] proved that the contribution of terms $\sim 1/R$ becomes essential at radiuses less than 5-50nm at the reasonable values of surface stress tensor $|\mu| = 5-50\,\text{N/m}$ [34]. At positive characteristic constants size-induced phase transition exists, while the temperature enhancement is possible at their negative values. The inequality $T_{cm}^*/T_{ce}^* < 1$ is typical for to the majority of bulk multiferroic materials such as BiFeO$_3$, Pb(Fe,Nb)O$_3$, Eu(Ba,Ti)O$_3$ [15]. However it is not excluded $T_{cm}(R)/T_{ce}(R) < 1$ for nanorods of definite radius $R$ allowing for the considered size effects as shown in Figs.5(a-d).

Also it is important for further consideration that characteristic parameters $R_{Q,Z}$ (as well as parameters $\rho_{A,B}$), which determine the size-induced transition temperature shift in accordance with Eqs.(16), and parameters $R_{ij}$, which determine the magnetoelectric coupling coefficients size effects in accordance with Eq.(11), depend on the different material constants (e.g. piezo-constants), have different numerical values and thus should be tuned independently. In fact we deal with double-scale size effect.

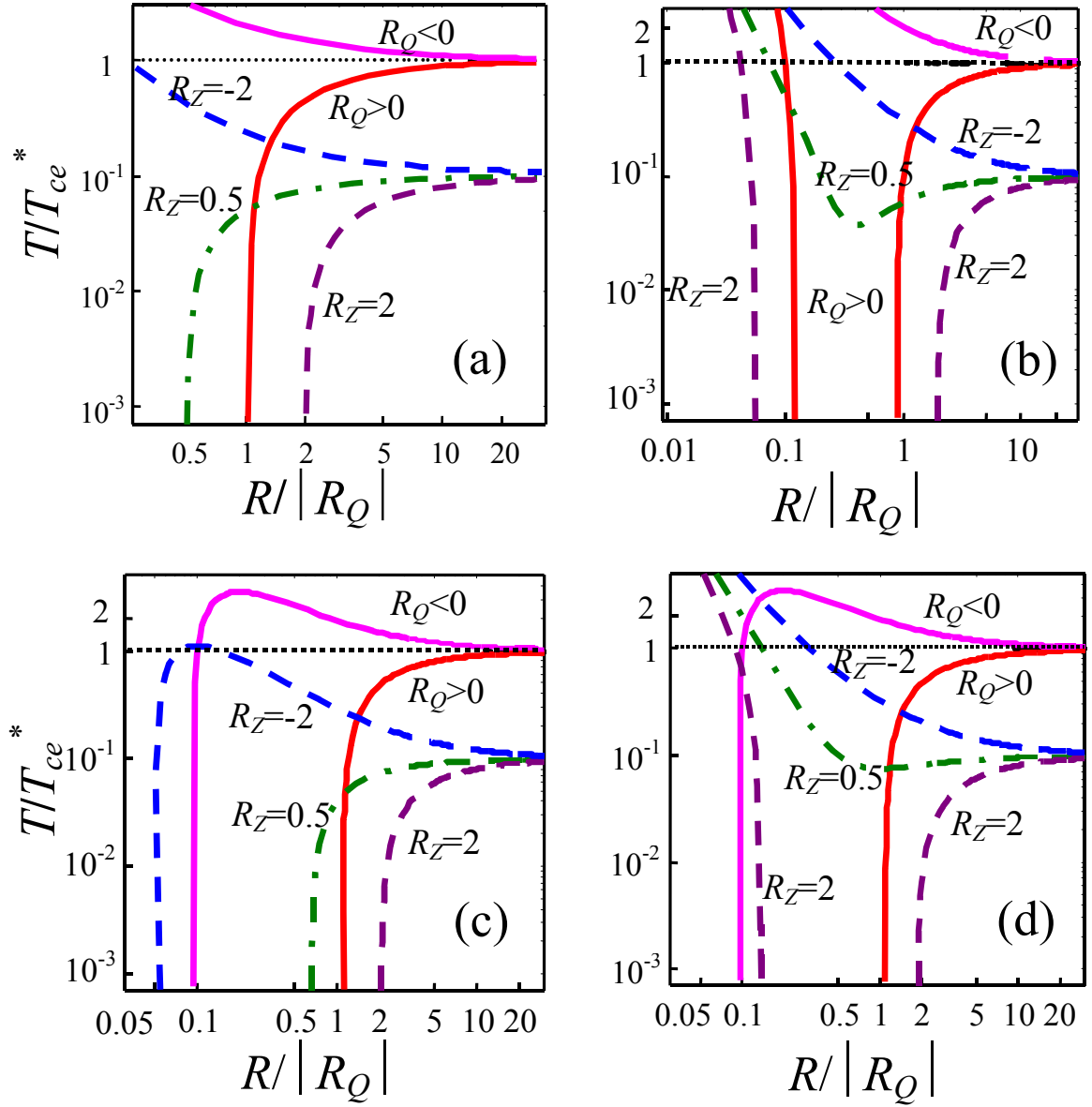

Fig.5. (Color online) Schematic dependence of the temperatures $T_{ce}(R)$ (solid curves) and $T_{cm}(R)$ (dashed curves for $R_Z/|R_Q|=\pm 2$ and dash-dotted curves for $R_Z/|R_Q|=0.5$) via the nanorod radius $R/|R_Q|$ for $T^*_{cm}/T^*_{ce}=0.1$ and (a) $\rho_{A,B}=0$; (b) $\rho_{A,B}=0.1\cdot R_Q^2$; (c) $\rho_{A,B}=-0.1\cdot R_Q^2$; (d) $\rho_A=-0.1\cdot R_Q^2$, $\rho_B=0.25\cdot R_Q^2$.

*6.2 Size effect on phase diagrams with ME coupling*

Below we mainly considered the most widely spread case of existence of the squared ME effect only (i.e. piezoeffect is absent and so $\gamma_{11}=0$, $\gamma_{12}=0$, $\gamma_{21}=0$, $\gamma_{22}\neq 0$, $H_p=E_p=0$, $T^*_{cm}=T^b_{cm}$, $T^*_{ce}=T^b_{ce}$) to show that even in this case its influence allowing for size effects becomes very strong. Numerical analyses of Eqs.(13-15) proved that quadratic magnetoelectric coupling term $\gamma_{22}M_1^2P_3^2$ acts as ferroelectricity and ferromagnetism suppressing or enhancing factor depending on $\gamma_{22}$ sign. Positive $\gamma_{22}$ values decrease the phase transition temperature, while the negative ones increase it. Hereinafter we consider positive $\gamma_{22}$ values.

*6.2.1. Phase diagrams without external fields*

For the case when magnetic and electric fields are absent ($H_0 = 0$, $E_0 = 0$), analysis of the free energy (12) is essentially simplified. In the Table I we summarized general conditions for the stability and existence of the four different phases, namely paraphase ($P = 0$, $M = 0$), ferroelectric ($P \neq 0$, $M = 0$), ferromagnetic ($P = 0$, $M \neq 0$) and mixed ferroelectric - ferromagnetic phase (secondary ferroic phase $P \neq 0$, $M \neq 0$) denoted as PP, FE, FM and FEM respectively.

**Table I**. The conditions of phases existence and stability in the absence of external fields

| Phase | Conditions |
|---|---|
| PP | $\alpha_1 > 0$, $\beta_1 > 0$ |
| FE | $\alpha_1 < 0$, $\alpha_1\gamma_{22} - 2\beta_1\alpha_{11} < 0$ |
| FM | $\beta_1 < 0$, $\beta_1\gamma_{22} - 2\alpha_1\beta_{11} < 0$ |
| FEM | $\alpha_1\gamma_{22} - 2\beta_1\alpha_{11} > 0$, $\beta_1\gamma_{22} - 2\alpha_1\beta_{11} > 0$, $4\alpha_{11}\beta_{11} - \gamma_{22}^2 > 0$ (Secondary Ferroic Phase) |

Since the free energy (12) renormalized coefficients (6), (7), (9) depend on the rod radius, Table I allows to construct phase diagrams in different coordinates. As one can see from the table, there are no ranges of coexistence between PP and FE, FM phases, or FEM and FE, FM, PP phases. At the same time there is some possibility for FE and FM phases to coexist. In this case the phase transition between FE and FM would be of first order. The phase transition between other phases (PP with FE or FM; or FEM with FE or FM or PP) is of the second order. The conditions of the phase transitions are summarized in the Table II.

**Table II**. The boundaries between different phases (if any) in the absence of external fields

| Transition | Order | Condition |
|---|---|---|

Giant magnetoelectric effect induced by intrinsic surface stress in ferroic nanorods

| PP - FE | II | $\alpha_1 = 0$ (i.e. $T = T_{ce}(R)$) |
|---|---|---|
| PP - FM | II | $\beta_1 = 0$ (i.e. $T = T_{cm}(R)$) |
| FE - FM | I | $\alpha_1^2/\alpha_{11} = \beta_1^2/\beta_{11}$ |
| FE - FEM | II | $\alpha_1\gamma_{22} = 2\beta_1\alpha_{11}$ (i.e. $T = T_{cm}(R) - \dfrac{\gamma_{22}(R)}{\beta_T}P_3^2(T,R)$) |
| FM - FEM | II | $\beta_1\gamma_{22} = 2\alpha_1\beta_{11}$ (i.e. $T = T_{ce}(R) - \dfrac{\gamma_{22}(R)}{\alpha_T}M_1^2(T,R)$) |

Phase diagrams of the considered system for different values of ratio $T_{cm}^b/T_{ce}^b$ and $R_Z/R_Q$ and small bulk value $\Gamma_{22}^b = 10^{-2}$ are presented in Fig. 6.

Hereinafter the ferroelectric to ferromagnetic energy ratio $W = \left(\beta_T T_{cm}^b\sqrt{\alpha_{11}}\right)/\left(\alpha_T T_{ce}^b\sqrt{\beta_{11}}\right)$ and dimensionless quadratic magnetoelectric coupling coefficient $\Gamma_{22}(R) = \gamma_{22}(R)/\sqrt{4\alpha_{11}\beta_{11}}$ are introduced. Also we introduce normalized electric and magnetic fields $\widetilde{E} = E_0/E_C$ and $\widetilde{H} = H_0/H_C$ respectively, where the values $E_C = 2\alpha_T T_{ce}^b P_S$ and $H_C = 2\beta_T T_{cm}^b M_S$ are proportional to the thermodynamic coercive fields.

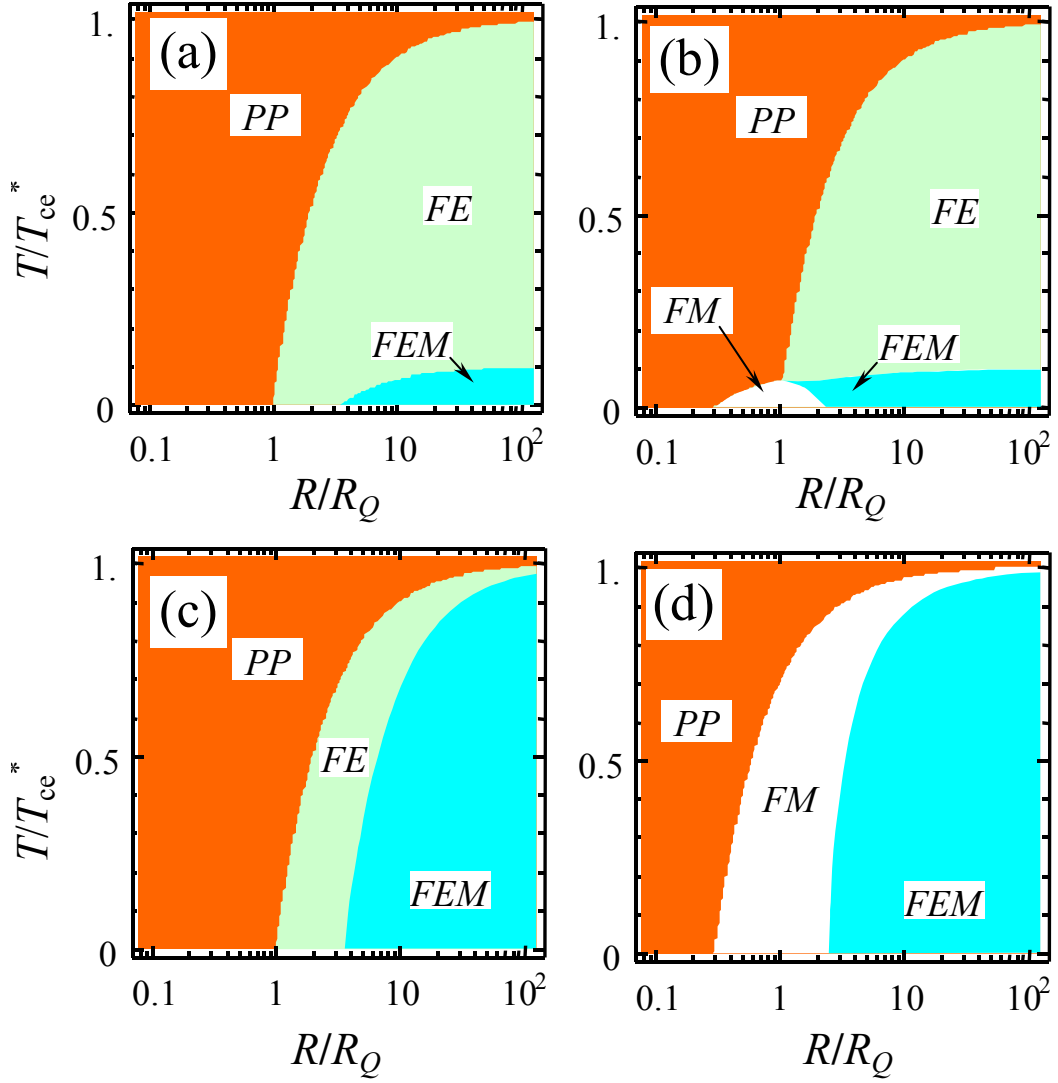

Fig. 6. (Color online). Phase diagrams for parameters $\Gamma_{22}^b = 10^{-2}$, $W = 1.5$, $R_Q > 0$, $R_{22}/R_Q = 100$, $\rho_{A,B} = 0$, $\tilde{H} = 0$, $\tilde{E} = 0$ and (a) $R_Z/R_Q = 3$, $T_{cm}^b/T_{ce}^b = 0.1$; (b) $R_Z/R_Q = 0.3$, $T_{cm}^b/T_{ce}^b = 0.1$; (c) $R_Z/R_Q = 3$, $T_{cm}^b/T_{ce}^b = 1$; (d) $R_Z/R_Q = 0.3$, $T_{cm}^b/T_{ce}^b = 1$.

Phase diagrams of the considered system for different values of ratio $T_{cm}^b/T_{ce}^b$, different sign of $R_Q$ and $R_Z > |R_Q|$ and large bulk value $\Gamma_{22}^b = 0.5$ are presented in Fig. 7.

Giant magnetoelectric effect induced by intrinsic surface stress in ferroic nanorods

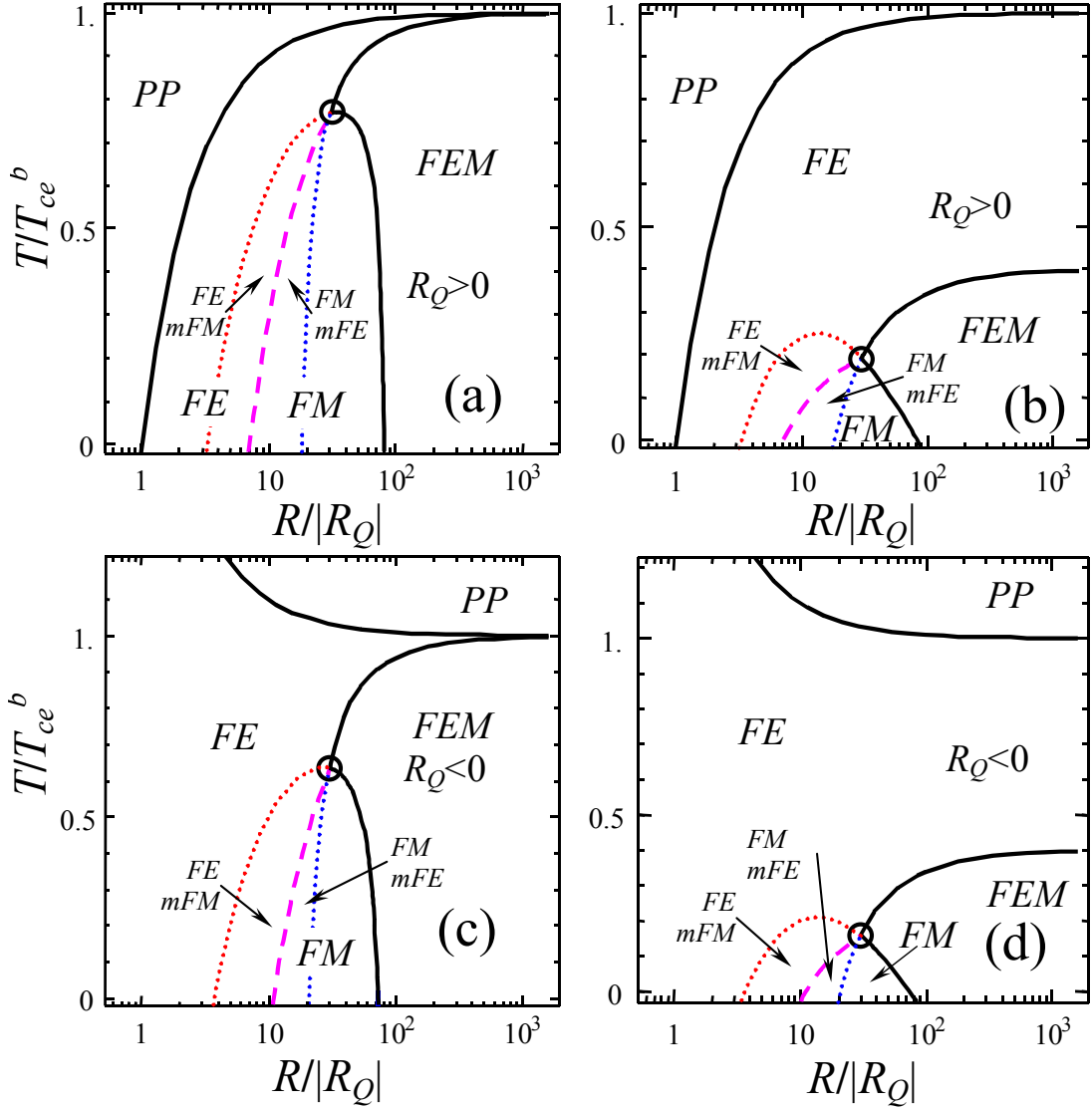

Fig. 7. (Color online). Phase diagrams for parameters $\Gamma_{22}^b = 0.5$, $W = 1.5$, $R_Z/|R_Q| = 3$, $R_{22}/|R_Q| = 30$, $\rho_{A,B} = 0$, $\tilde{H} = 0$, $\tilde{E} = 0$ and (a) $R_Q > 0$, $T_{cm}^b/T_{ce}^b = 1$; (b) $R_Q > 0$, $T_{cm}^b/T_{ce}^b = 0.5$; (c) $R_Q < 0$, $T_{cm}^b/T_{ce}^b = 1$; (d) $R_Q < 0$, $T_{cm}^b/T_{ce}^b = 0.5$. Circles denote triple points. The second and the first order phase transitions are shown by solid and dashed lines, respectively. Dotted lines denote the limits of different phases stability.

Comparing Figs. 7a with 7c or 7b with 7d, one can see the change of $R_Q$ sign drastically changes the transition line between FE and PP phase, so that the transition temperature trend changes from the decrease to the increase with decreasing radius. At the same time, other transitions dependences on radius qualitatively remain the same.

It should be noted that the regions of FE and FM phase coexistence are situated between dotted lines in Fig. 7. In this case equilibrium phase transition between these phases will be of

Giant magnetoelectric effect induced by intrinsic surface stress in ferroic nanorods first order, when the free energy values for coexisting phases are equal. Thus, the region of coexistence that corresponds to the secondary ferroic appearance is divided into sub-regions where one of phases has the deeper minima and is absolutely stable, while the other is metastable. Hereinafter mFE, mFM and mFEM denote metastable FE, FM and FEM phases respectively.

To illustrate the triple point properties and compare this state with the other ones, we presented free energy contours for triple point in Fig. 8a along with contours FE, FEM and FM phases in Figs. 8b, c and d respectively for the parameters corresponding to Fig. 7a.

It is seen from Fig. 8a that the triple point state is characterized by a continuous set of order parameters, since any point ($P$, $M$) lying on the dotted circle in Fig.7a corresponds to the free energy minimum.

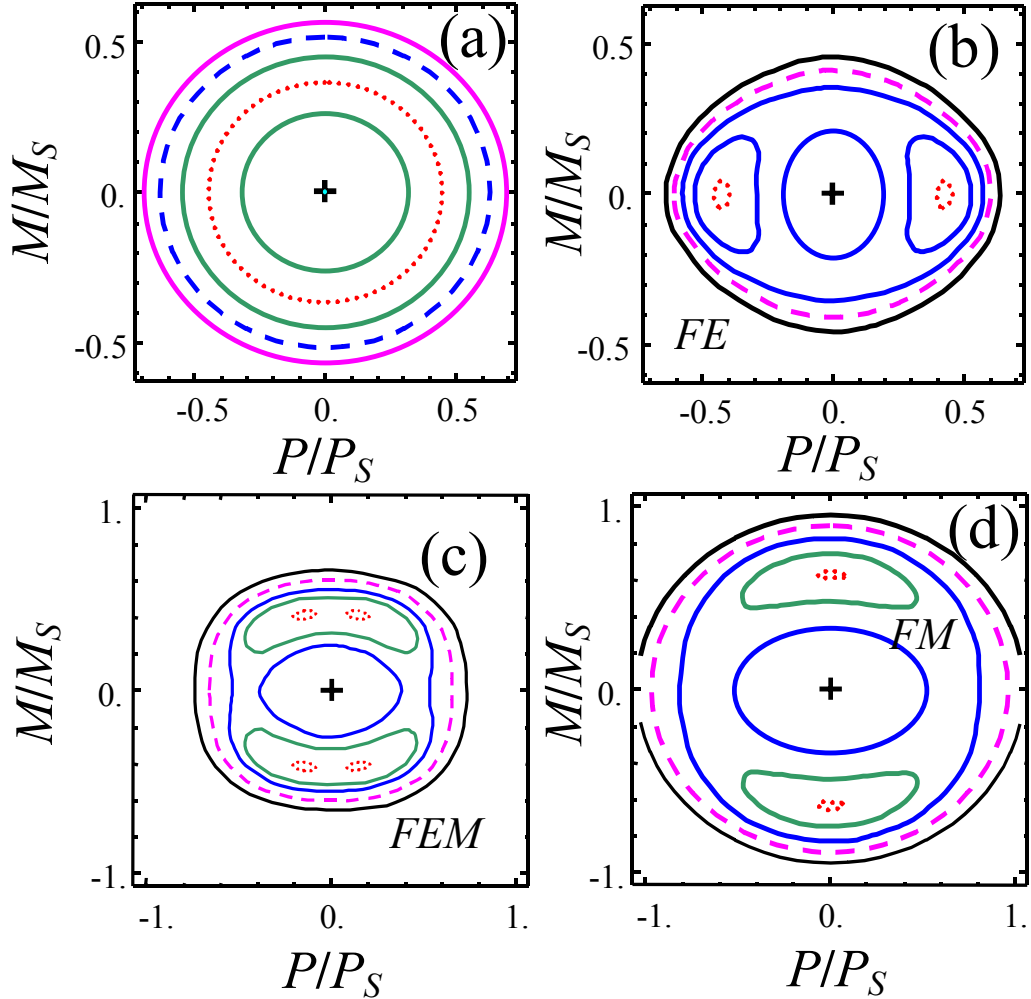

Fig. 8. (Color online). Free energy contours for parameters $\Gamma_{22}^b = 0.5$, $W = 1.5$, $R_Z/|R_Q| = 3$, $R_{22}/R_Q = 30$ ($R_Q > 0$), $T_{cm}^b/T_{ce}^b = 1$, $\rho_{A,B} = 0$, $\tilde{H} = 0$, $\tilde{E} = 0$ and (a) $T/T_{ce}^b = 0.77$, $R/R_Q = 30$

Giant magnetoelectric effect induced by intrinsic surface stress in ferroic nanorods

(triple point); (b) $T/T_{ce}^b = 0.77$, $R/R_Q = 20$ (ferroelectric phase); (c) $T/T_{ce}^b = 0.77$, $R/R_Q = 60$ (secondary ferroic phase); (d) $T/T_{ce}^b = 0.5$, $R/R_Q = 30$ (ferromagnetic phase). Crosses denote maxima positions. Dotted and dashed contours correspond to the free energy minima and zero values respectively.

In order to consider the case of $T_{cm}^b = 0$, $T_{ce}^b = 0$ (i.e. when bulk material has no ferroic properties), one should use Eqs.(16c,d) and introduce $W = \sqrt{\alpha_{11}/\beta_{11}}$. Phase diagrams for the case of $T_{cm}^b = 0$, $T_{ce}^b = 0$ and different values of $R_Z$, $\gamma_{22}^b$ and $\beta_T/\alpha_T$ is shown in Figs. 9 for the typical case $\beta_T/\alpha_T = 10^2$ (for the majority of ferroic materials $\beta_T/\alpha_T \gg 1$). In this case ferroic phase could exist only at small radii, the transition temperatures increase with radius decrease. PP phase transit into the phase with higher transition temperature, FE or FM phase. As it is seen from Figs. 9, the region of FEM phase existence is wider for the case $\gamma_{22} < 0$ than for $\gamma_{22} > 0$ (compare panels (a) and (b), (c) and (d)) as anticipated.

Phase diagrams corresponding to the cases $\beta_T/\alpha_T = 1$ and $\beta_T/\alpha_T = 10^{-2}$ are shown in Appendix C.

Giant magnetoelectric effect induced by intrinsic surface stress in ferroic nanorods

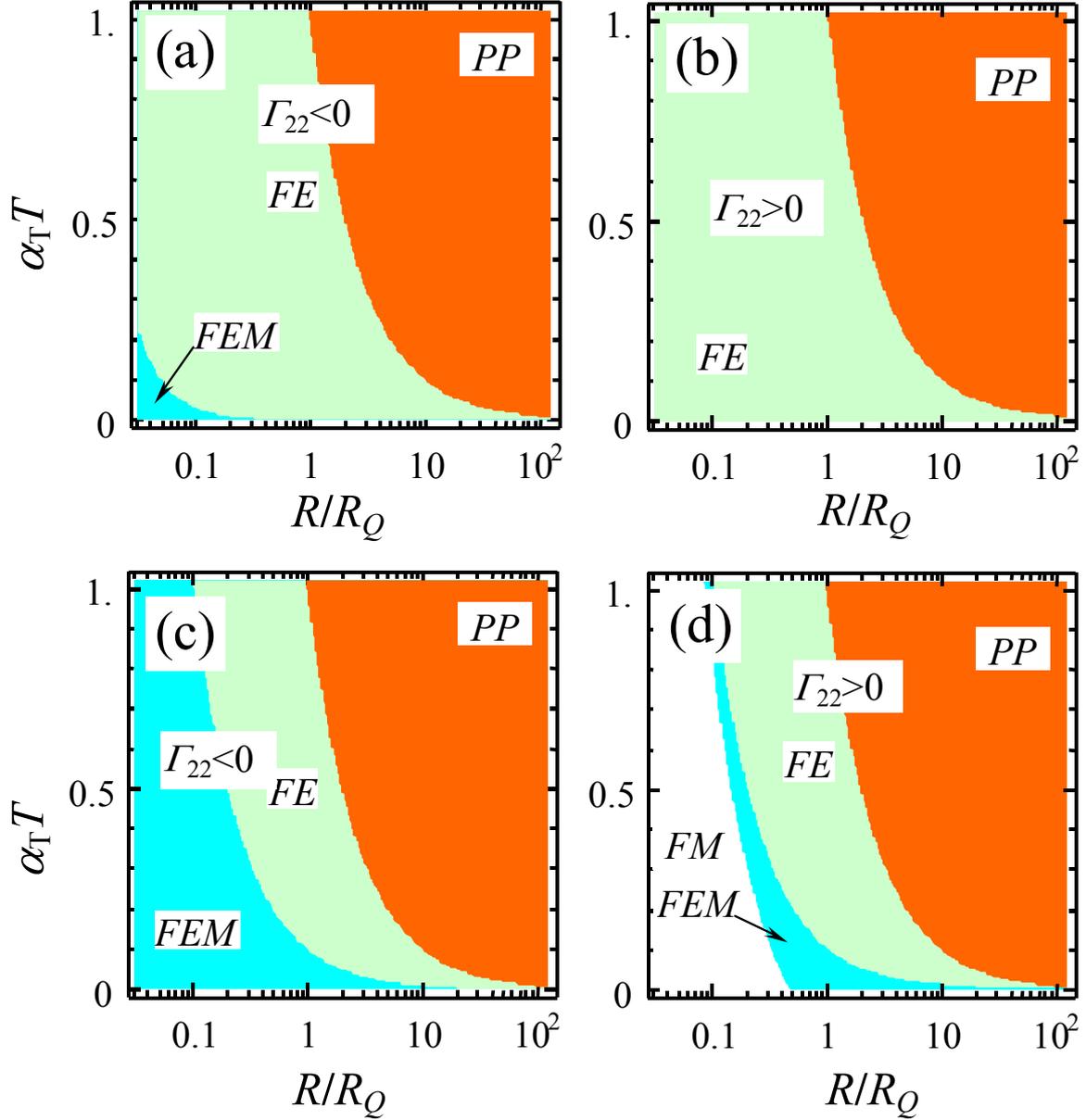

Fig. 9. (Color online). Phase diagrams for following parameters $\beta_T/\alpha_T = 10^2$, $T_{cm}^b = 0$, $T_{ce}^b = 0$, $W = 1.5$, $R_Q > 0$, $R_{22}/R_Q = 30$, $\rho_{A,B} = 0$, $\tilde{H} = 0$, $\tilde{E} = 0$ and (a) $\Gamma_{22}^b = -10^{-3}$, $R_Z/R_Q = 0.1$; (b) $\Gamma_{22}^b = 10^{-3}$, $R_Z/R_Q = 0.1$; (c) $\Gamma_{22}^b = -10^{-3}$, $R_Z/R_Q = 10$; (d) $\Gamma_{22}^b = 10^{-3}$, $R_Z/R_Q = 10$.

### 6.2.2. Phase diagrams under external fields

The application of the electric or magnetic field induces corresponding order parameter (polarization or magnetization), so that formally paraphase could not be introduced (although the reversibility of the state could help to distinguish it from paraphase). At the same time mixed phase FEM could exist. Typically two limiting cases $E_0 = 0$ and $H_0 = 0$ have been considered, namely:

Giant magnetoelectric effect induced by intrinsic surface stress in ferroic nanorods

(a) In the absence of electric field ($E_0 = 0$) the system under magnetic field could be either in FM or in FEM phases. For the case $4\alpha_{11}\beta_{11} - \gamma_{22}^2 > 0$, $\beta_1\gamma_{22} - 2\alpha_1\beta_{11} > 0$, $\alpha_1 < 0$ FEM is stable only for the magnetic fields $H_0 < H_I$, where $H_I$ is the first critical magnetic field. For the fields above critical value $H_0 > H_I$ FEM phase transforms into FM one. The critical value of the field along with the aforementioned necessary conditions of its existence can be written as

$$H_I = 2\sqrt{-\frac{\alpha_1}{\gamma_{22}}\left(\beta_1 - 2\frac{\alpha_1}{\gamma_{22}}\beta_{11}\right)}. \tag{17}$$

For the case $4\alpha_{11}\beta_{11} - \gamma_{22}^2 < 0$ FEM phase is stable only at $H_0 > H_{II}$, where the second critical value of the field can be written as:

$$H_{II} = \frac{2}{3}\sqrt{\frac{\alpha_1\gamma_{22} - 2\beta_1\alpha_{11}}{3(4\alpha_{11}\beta_{11} - \gamma_{22}^2)}}\left(2\beta_1 - \frac{\gamma_{22}}{\alpha_{11}}\alpha_1\right). \tag{18}$$

In the region where first order phase transition takes place both expressions (17-18) make sense and there is hysteresis for the fields $H_I < H_0 < H_{II}$.

(b) In the absence of magnetic field ($H_0 = 0$) the system under electric field could be in either in the FE or FEM phases. For the case $4\alpha_{11}\beta_{11} - \gamma_{22}^2 > 0$, $\alpha_1\gamma_{22} - 2\beta_1\alpha_{11} > 0$, $\beta_1 < 0$ FEM is stable only for the fields $E_0 < E_I$, for the field above critical value $E_I$ FEM phase transforms into FE one:

$$E_I = 2\sqrt{-\frac{\beta_1}{\gamma_{22}}\left(\alpha_1 - 2\frac{\beta_1}{\gamma_{22}}\alpha_{11}\right)}. \tag{19}$$

For the case $4\alpha_{11}\beta_{11} - \gamma_{22}^2 < 0$ FEM phase is stable only at $E_0 > E_{II}$, where critical value of the field can be written as:

$$E_{II} = \frac{2}{3}\sqrt{\frac{\beta_1\gamma_{22} - 2\alpha_1\beta_{11}}{3(4\alpha_{11}\beta_{11} - \gamma_{22}^2)}}\left(2\alpha_1 - \frac{\gamma_{22}}{\beta_{11}}\beta_1\right). \tag{20}$$

For the region where first order phase transition takes place, both expressions (19-20) make sense and there is hysteresis for the fields $E_I < E_0 < E_{II}$.

Using the expressions (17)-(20), one can consider how the typical zero-fields phase diagram from Fig. 6b and 6d changes under the presence of external fields as presented in Figs. 10a, c and Figs. 10b, d for the cases of electric and magnetic field respectively. It is seen that under the field increase in the region of FEM phase existence is narrowed (compare Figs. 10a with 10b or Figs. 10c with 10d), e.g. corresponding transitions/stability limits temperatures shift to lower values.

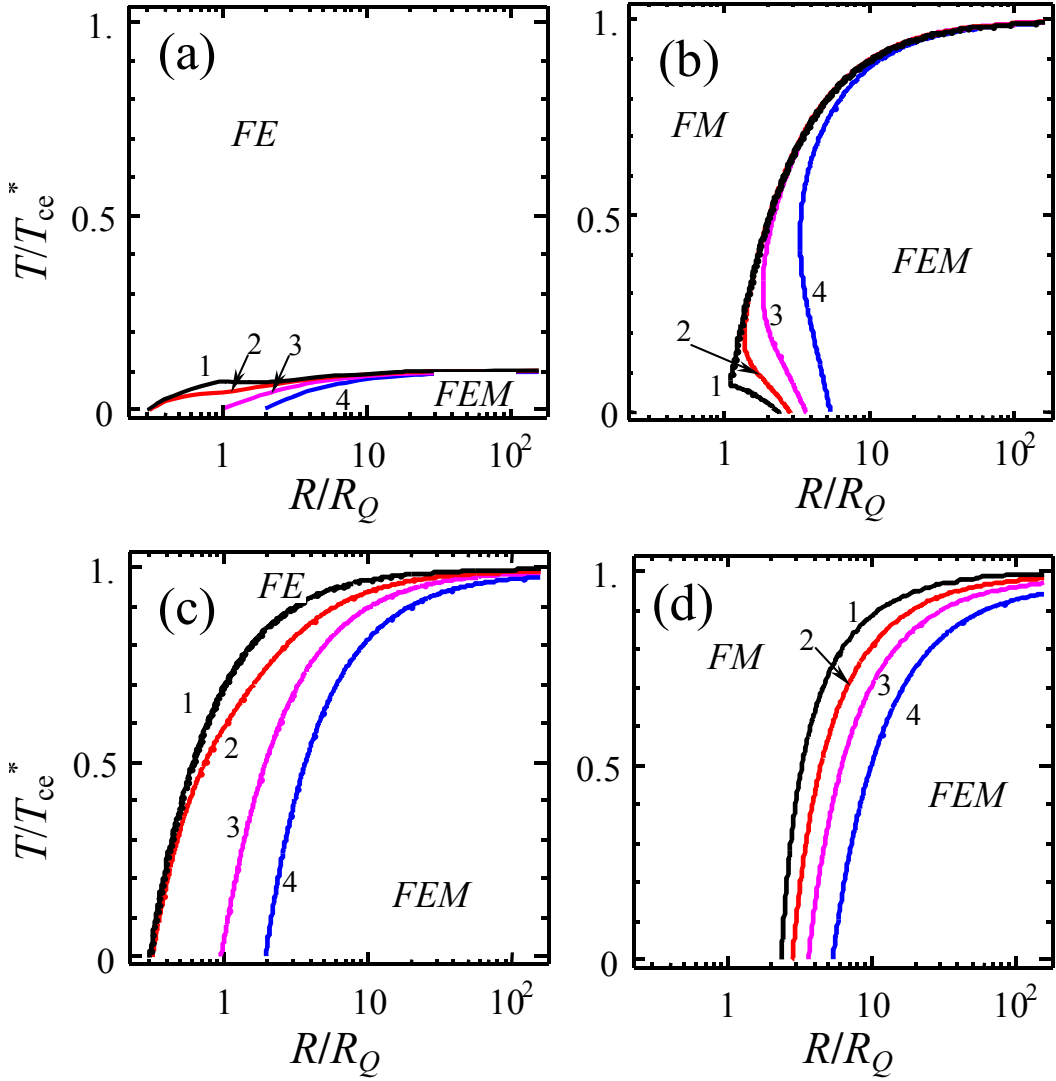

Fig. 10. (Color online). Phase diagrams for parameters $\Gamma_{22}^b = 10^{-2}$, $W = 1.5$, $R_Q > 0$, $R_{22}/R_Q = 100$, $R_Z/R_Q = 0.3$, $\rho_{A,B} = 0$. (a) $T_{cm}^b/T_{ce}^b = 0.1$, $\tilde{H} = 0$, $\tilde{E} = 10^{-4}, 0.3, 1, 3$ (curves 1, 2, 3, 4); (b) $T_{cm}^b/T_{ce}^b = 0.1$, $\tilde{H} = 10^{-4}, 0.3, 1, 3$, $\tilde{E} = 0$ (curves 1, 2, 3, 4); (c) $T_{cm}^b/T_{ce}^b = 1$, $\tilde{H} = 0$, $\tilde{E} = 10^{-4}, 0.3, 1, 3$ (curves 1, 2, 3, 4); (d) $T_{cm}^b/T_{ce}^b = 1$, $\tilde{H} = 10^{-4}, 0.3, 1, 3$, $\tilde{E} = 0$ (curves 1, 2, 3, 4).

The changes of zero-fields phase diagram from Fig. 7a under the presence of external fields are presented in Figs. 11a-b and Figs. 11c-d for the cases of magnetic and electric field respectively. It is seen that under the field increase the region of FEM phase existence is narrowed slightly (compare Figs. 11a with 11b or Figs. 11c with 11d), i.e. corresponding transitions/stability limits temperatures shift to lower values.

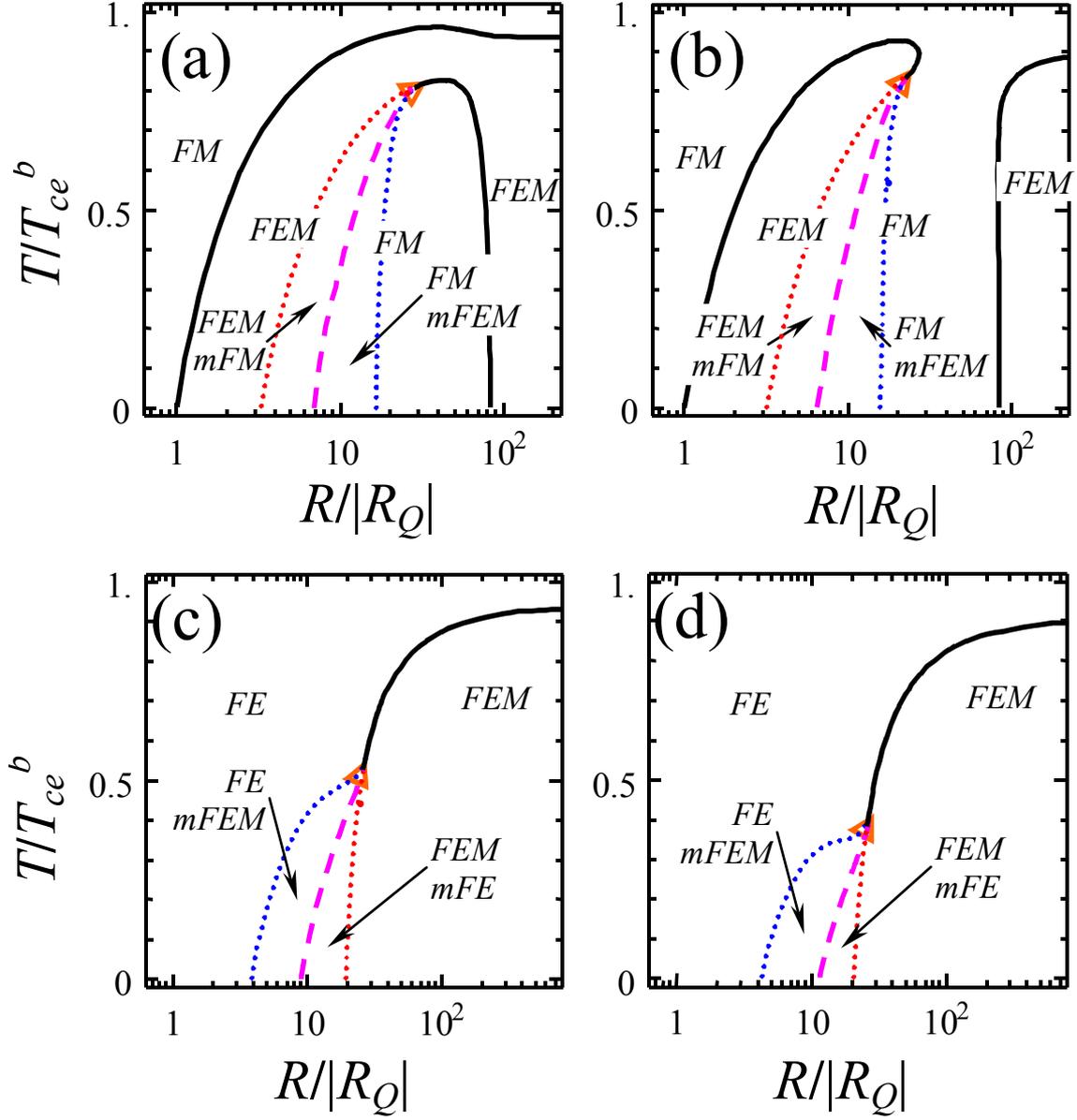

Fig. 11 (Color online). Phase diagrams for parameters $\Gamma^b_{22} = 0.5$, $W = 1.5$, $R_Z/R_Q = 3$, $R_{22}/R_Q = 30$ ($R_Q > 0$), $T^b_{cm}/T^b_{ce} = 1$, $\rho_{A,B} = 0$ and (a) $\widetilde{H} = 0.005$, $\widetilde{E} = 0$; (b) $\widetilde{H} = 0.01$, $\widetilde{E} = 0$; (c) $\widetilde{H} = 0$, $\widetilde{E} = 0.05$; (d) $\widetilde{H} = 0$, $\widetilde{E} = 0.1$. Triangles denote tricritical points. The second- and first-order phase transitions are shown by solid and dashed lines, respectively. Dotted lines denote the limits of different phases stability.

Giant magnetoelectric effect induced by intrinsic surface stress in ferroic nanorods

**Conclusion**

- We predict the effects related with the renormalization of the magnetoelectric coupling coefficients caused by intrinsic surface stress in ferroic nanorods. The linear coupling coefficient is radius independent, whereas the quadratic ones include terms inversely proportional to the nanorod radius and thus strongly increase with radius decrease.

- The renormalized magnetoelectric effect increases the relative dielectric tunability 3-50 times. At small magnetic field the magnetoelectric tunability increases under the magnetic field increase up to 2-5 times. A jump of the relative dielectric tunability is related with ferromagnetic phase transition shifted by the ME coupling. The jump height increases with magnetic field increase. Giant tunability (more than 50 times) appearance at high enough magnetic fields is caused by the ME coupling-induced ferroelectric-paraelectric phase transition taken place at positive coupling coefficient.

- The quadratic magnetoelectric coupling dramatically changes the phase diagrams of ferroic nanorods with a decrease of the radius. The method to construct different phase diagrams by changing the nanorod radii is proposed. A ME coupling-induced ferroelectric-paraelectric phase transition (absent in the bulk) takes place at high enough magnetic fields. The transition appeared at positive coupling coefficients, when the magnetoelectric effect suppresses theferroelectric phase. Also the second order phase transition may become a first one. The triple point appears at zero external electric and magnetic fields tricritical points appear also under the external fields.

- The intrinsic surface stress under the curved nanorod surface shifts the ferroelectric and ferromagnetic transition temperatures. The corresponding transition temperature shift (unrelated with ME effect) and negative magnetoelectric coupling may induce ferromagnetism in small nanorods absent in the bulk material. Similar mechanism could explain recently observed room temperature ferromagnetism in small nanoparticles of non-magnetic oxides.

- Under the presence of piezomagnetic and piezoelectric effects, the intrinsic surface stress induces built-in magnetic and electric fields correspondingly. Built-in fields are inversely proportional to the nanorod radius. The fields smear magnetic, dielectric and magnetoelectric susceptibility temperature maxima, and increase their values in the paraphase. Built-in fields may overcome the coercive fields and thus essentially increase dielectric tunability even in the absence of external fields.

**Appendix A. Estimations**

1). Let us estimate the coefficient $A_{33}$ involved in (1). This could be done from the jump of elastic compliance $s_{11}$ in the point of the bulk ferroelectric phase transition at $T = T_{ce}$ [35], since $\Delta s_{11}^{P,E} = \left(s_{11}^{P,E} - s_{11}\right) \sim P_3^2$, namely $\Delta s_{12}^{P,E} = A_{33} P_3^2$. In accordance with data for BaTiO$_3$ in Fig.4.3 of Ref.[36] one obtains that $\Delta s_{11}^{P} = -1.6 \cdot 10^{-12}$ Pa and $\Delta s_{11}^{E} = 6 \cdot 10^{-12}$ Pa; whereas $P_3^2 = 0.2\,\text{C/m}^2$ at 100$^0$C. Thus $A_{33}^{P} = -4 \cdot 10^{-11}\,\text{m}^4/\text{C}^2$ Pa and $A_{33}^{E} = +15 \cdot 10^{-11}\,\text{m}^4/\text{C}^2$Pa.

Then let us estimate the relative shift of the magnetoelectric coefficient $\gamma_{12}$ determined by the radius $R_{12} = 2\mu \dfrac{s_{12} A_{33}^{P,E}}{s_{11} Q_{11}}$. Using typical parameters $Q_{11} = 0.11\,\text{m}^2/\text{C}^2$, the surface stress $\mu = (5-50)\,\text{N/m}$ and $s_{12}/s_{11} = -0.3$ (since $s_{12} < 0$ for cubic perovskites) we obtained $R_{12}^{P} = (1.2-12)\,\text{nm}$ and $R_{12}^{E} = -(4.5-45)\,\text{nm}$.

2). Let us estimate the coefficient $B_{33}$ involved in (1) and magnetoelectric coefficient $\gamma_{21}$ determined by the radius $R_{21} = 2\mu \dfrac{s_{12} B_{33}}{s_{11} Z_{11}}$. Using the relation ship $\Delta s_{12}^{M,H} = B_{33} M_1^2$ and typical values $M_1 \sim 1 \div 0.3$ T (Tesla), $M_1^2 \sim 1 \div 0.1$ ($M_1^2 \sim 10 \div 0.1$) T$^2$, $\Delta s_{12} \sim 10^{-12}$ Pa one obtains that $B_{33} = (1 \div 0.1) \cdot 10^{-11}$ Pa/T$^2$ Si units. The estimation for $\mu \cong 5 \cdot 10^4$ din/cm, bulk magnetostriction coefficients $Z_{ij} \sim 10^{-11} - 10^{-9}\,cm^3/erg$ typical for rare-earth alloys [8] and $s_{12}/s_{11} = -0.3$ leads to the values $|R_{21}| = (5-50)\,\text{nm}$.

**Appendix B. Generalized susceptibilities calculations.**

Using Eq. (14), the following relations for full differentials can be written:

$$\begin{cases} dE_0 = \dfrac{\partial^2 \widetilde{g}}{\partial P_3^2} dP_3 + \dfrac{\partial^2 \widetilde{g}}{\partial P_3 \partial M_1} dM_1, \\ dH_0 = \dfrac{\partial^2 \widetilde{g}}{\partial P_3 \partial M_1} dP_3 + \dfrac{\partial^2 \widetilde{g}}{\partial M_1^2} dM_1, \end{cases} \quad (\text{B.1})$$

where $\dfrac{\partial^2 \widetilde{g}}{\partial P_3^2} = 2\alpha_1 + 2\gamma_{12} M_1 + 2\gamma_{22} M_1^2 + 12\alpha_{11} P_3^2$, $\dfrac{\partial^2 \widetilde{g}}{\partial M_1^2} = 2\beta_1 + 2\gamma_{21} P_3 + 2\gamma_{22} P_3^2 + 12\beta_{11} M_1^2$ and $\dfrac{\partial^2 \widetilde{g}}{\partial P_3 \partial M_1} = \gamma_{11} + 2\gamma_{12} P_3 + 2\gamma_{21} M_1 + 4\gamma_{22} M_1 P_3$. After elementary transformations electric and magnetic susceptibilities can be found from the system (B.1) as the inverse matrix in the form of Eqs.(15).

In the paramagnetic phase ($M_1 \to 0$ and infinitely small magnetic fields) expressions (14) acquires the form:

$$\chi_E = \frac{1}{2(\alpha_1 + 6\alpha_{11}P_3^2) - (\gamma_{11} + 2\gamma_{12}P_3)^2 / (2(\beta_1 + \gamma_{21}P_3 + \gamma_{22}P_3^2))}, \tag{B.2a}$$

$$\chi_M = \frac{1}{2(\beta_1 + \gamma_{21}P_3 + \gamma_{22}P_3^2) - (\gamma_{11} + 2\gamma_{12}P_3)^2 / (2\alpha_1 + 12\alpha_{11}P_3^2)}, \tag{B.2b}$$

$$\chi_{ME} = \frac{-(\gamma_{11} + 2\gamma_{12}P_3)}{4(\alpha_1 + 6\alpha_{11}P_3^2)(\beta_1 + \gamma_{21}P_3 + \gamma_{22}P_3^2) - (\gamma_{11} + 2\gamma_{12}P_3)^2}. \tag{B.2c}$$

In the paramagnetic and paraelectric phase ($M_1 \to 0$ and $P_3 \to 0$) at infinitely small magnetic and electric fields one obtains from Eqs.(B.2) that

$$\chi_E = \frac{2\beta_1}{4\alpha_1\beta_1 - \gamma_{11}^2}, \quad \chi_M = \frac{2\alpha_1}{4\alpha_1\beta_1 - \gamma_{11}^2}, \quad \chi_{ME} = \frac{-\gamma_{11}}{4\alpha_1\beta_1 - \gamma_{11}^2}. \tag{B.3}$$

It is follows from Eq.(B.3) that magnetoelectric coupling coefficient divergence corresponds to the condition $4\alpha_1\beta_1 - \gamma_{11}^2 = 0$.

Bulk susceptibilities in the case of quadratic magnetoelectric coupling:

$$\chi_{ME}^b = \frac{-\gamma_{22}M_1P_3}{(\alpha_1 + 6\alpha_{11}P_3^2 + \gamma_{22}M_1^2)(\beta_1 + 6\beta_{11}M_1^2 + \gamma_{22}P_3^2) + 4\gamma_{22}^2 M_1^2 P_3^2} \approx \frac{-\gamma_{22}M_1P_3}{\alpha_1\beta_1}, \tag{B.4a}$$

$$\chi_E^b = \frac{\beta_1 + \gamma_{22}P_3^2 + 6\beta_{11}M_1^2}{2(\alpha_1 + 6\alpha_{11}P_3^2 + \gamma_{22}M_1^2)(\beta_1 + 6\beta_{11}M_1^2 + \gamma_{22}P_3^2) + 8\gamma_{22}^2 M_1^2 P_3^2} \approx \frac{1}{\alpha_1}, \tag{B.4b}$$

$$\chi_M^b = \frac{\alpha_1 + \gamma_{22}M_1^2 + 6\alpha_{11}P_3^2}{2(\alpha_1 + 6\alpha_{11}P_3^2 + \gamma_{22}M_1^2)(\beta_1 + 6\beta_{11}M_1^2 + \gamma_{22}P_3^2) + 8\gamma_{22}^2 M_1^2 P_3^2} \approx \frac{1}{\beta_1}. \tag{B.4c}$$

**Appendix C. Phase diagrams.**

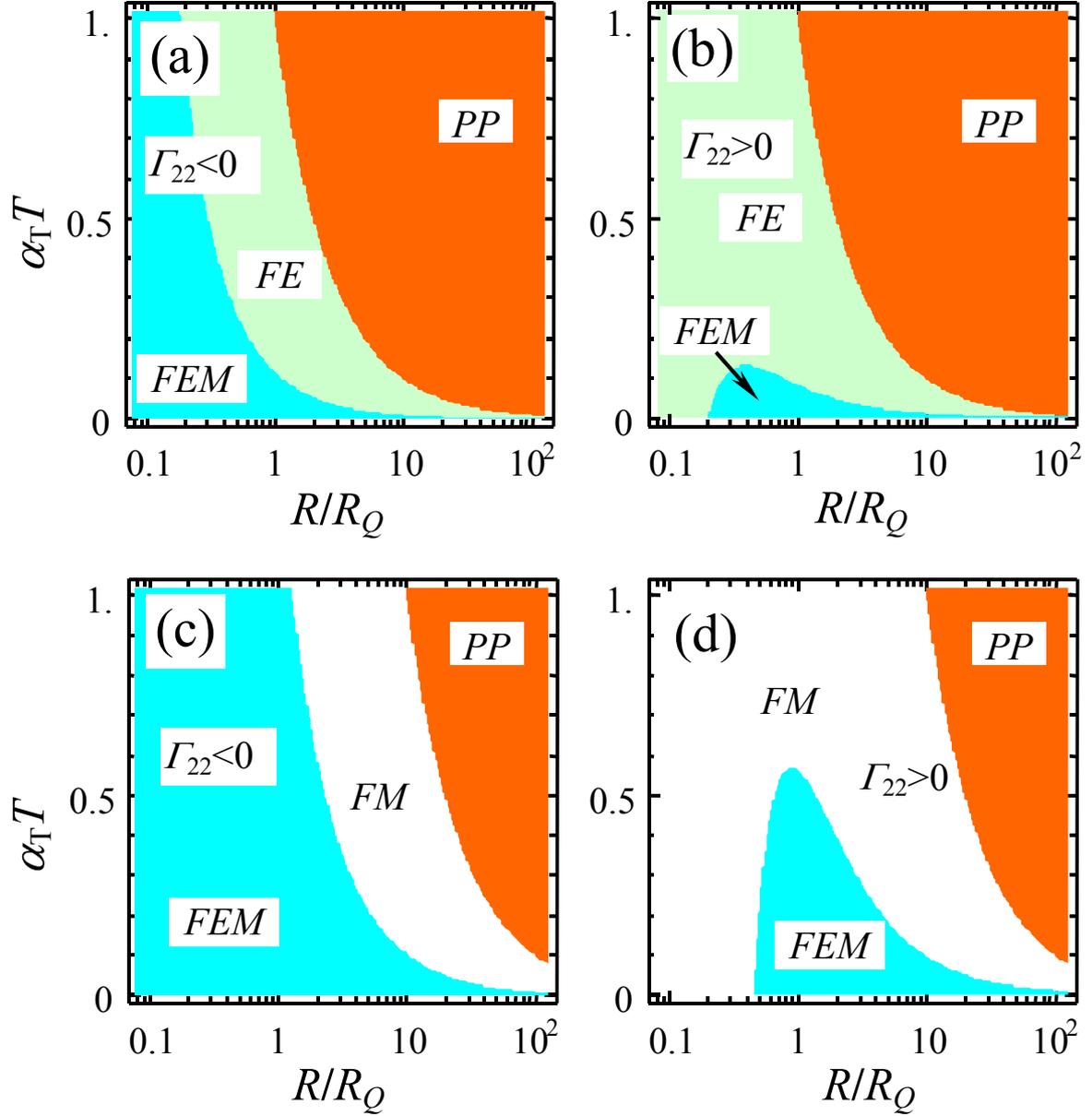

Fig. C1. (Color online). Phase diagrams for following parameters $\beta_T/\alpha_T = 1$, $T_{cm}^b = 0$, $T_{ce}^b = 0$, $W = 1.5$, $R_Q > 0$, $R_{22}/R_Q = 30$, $\rho_{A,B} = 0$, $\tilde{H} = 0$, $\tilde{E} = 0$ and (a) $\Gamma_{22}^b = -10^{-3}$, $R_Z/R_Q = 0.1$; (b) $\Gamma_{22}^b = 10^{-3}$, $R_Z/R_Q = 0.1$; (c) $\Gamma_{22}^b = -10^{-3}$, $R_Z/R_Q = 10$; (d) $\Gamma_{22}^b = 10^{-3}$, $R_Z/R_Q = 10$.

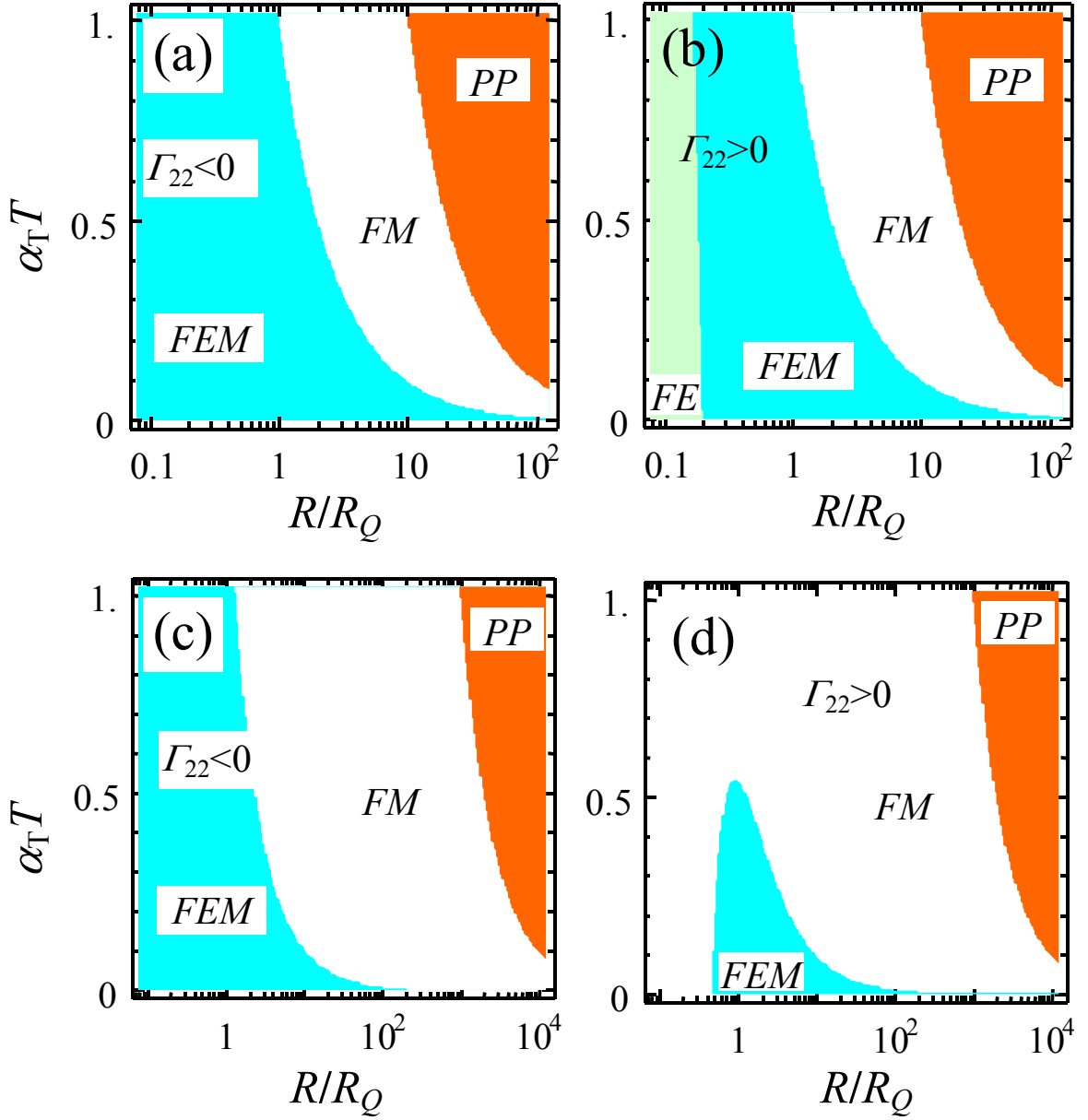

Fig. C2. (Color online). Phase diagrams for following parameters $\beta_T/\alpha_T = 10^{-2}$, $T^b_{cm} = 0$, $T^b_{ce} = 0$, $W = 1.5$, $R_Q > 0$, $R_{22}/R_Q = 30$, $\rho_{A,B} = 0$, $\tilde{H} = 0$, $\tilde{E} = 0$ and (a) $\Gamma^b_{22} = -10^{-3}$, $R_Z/R_Q = 0.1$; (b) $\Gamma^b_{22} = 10^{-3}$, $R_Z/R_Q = 0.1$; (c) $\Gamma^b_{22} = -10^{-3}$, $R_Z/R_Q = 10$; (d) $\Gamma^b_{22} = 10^{-3}$, $R_Z/R_Q = 10$.